\documentclass{LMCS}

\def\dOi{9(4:1)2013}
\lmcsheading%
{\dOi}
{1--32}
{}
{}
{Sep.~28, 2012}
{Oct.~\phantom01, 2013}
{}

\ACMCCS{[{\bf Software and its engineering}]: Software creation and management---Software verification and validation---Formal software verification}

 \usepackage{hyperref}

\usepackage{amsmath}
\usepackage{amssymb}
\usepackage{xcolor}
\usepackage{colortbl}%,hhline}
\usepackage[curve]{xypic}

\newcommand{\INRIA}{\mbox{\sc Inria}}

\newcommand{\BCG}{\mbox{\sc Bcg}}

\newcommand{\CADP}{\mbox{\sc Cadp}}

\newcommand{\CCS}{\mbox{\sc Ccs}}

\newcommand{\CSP}{\mbox{\sc Csp}}

\newcommand{\EXPOPEN}{\mbox{\sc Exp.Open}}
\newcommand{\EVALUATOR}{\mbox{\sc Evaluator}}

\newcommand{\LOTOS}{\mbox{\sc Lotos}}
\newcommand{\ELOTOS}{\mbox{\sc E-Lotos}}

\newcommand{\LOTOSNT}{\mbox{\sc Lotos NT}}

\newcommand{\LTS}{\mbox{\sc Lts}}
\newcommand{\LTSs}{\mbox{\sc Lts}s}

\newcommand{\mCRL}{\mbox{$\mu${\sc Crl}}}

\newcommand{\OPENCAESAR}{\mbox{\sc Open/C{\ae}sar}}

\newcommand{\SVL}{\mbox{\sc Svl}}

\newcommand{\dia}{\left<}
\newcommand{\mond}{\right>}
\newcommand{\brac}{\left[}
\newcommand{\ket}{\right]}
\newcommand{\sem}{\left[\!\left[}
\newcommand{\antic}{\right]\!\right]}

\newcommand{\Arrow}[1]{\stackrel{#1}{\rightarrow}}
\newcommand{\Lmu}{\mbox{$L{\mu}$}}
\newcommand{\Lmui}{\mbox{$L{\mu}_1$}}
\newcommand{\Lmuii}{\mbox{$L{\mu}_2$}}
\newcommand{\Lmun}{\mbox{$L{\mu}_n$}}
\newcommand{\LTL}{\mbox{\sc Ltl}}
\newcommand{\PDL}{\mbox{\sc Pdl}}
\newcommand{\PDLdelta}{\mbox{{\sc Pdl}-$\Delta$}}
\newcommand{\CTLstar}{\mbox{{\sc Ctl}$^{*}$}}
\newcommand{\Acyc}{\mbox{${{\rm A}_4}_{\it cyc}$}}
\newcommand{\holder}[2]{\overline{#1}^{#2}}

\newcommand{\COMMENT}[1]{}

\newcommand{\dblslash}{/\!\!/}
\newcommand{\trad}[3]{#1\,\dblslash^{#3}_{#2}}
\newcommand{\true}{\mathbf{t\!t}}
\newcommand{\false}{\mathbf{ff}}
\newcommand{\notop}{\neg}
\newcommand{\BES}{{\sc Bes}}
\newcommand{\lts}{\mathop{\mathsf{lts}}}
\newcommand{\trans}[1]{\smash{\stackrel{{#1}}{\longrightarrow}}}
\newcommand{\enc}{\mathop{\mathsf{enc}}}
\newcommand{\decs}{\mathop{\mathsf{dec}_{\mathsf{s}}}}
\newcommand{\dect}{\mathop{\mathsf{dec}_{\mathsf{t}}}}
\newcommand{\fv}{\mathop{\mathsf{fv}}}
\newcommand{\bv}{\mathop{\mathsf{bv}}}

\newcommand{\blocks}{\mathop{\mathsf{blocks}}}
\newcommand{\bl}{{\sf bl}}
\newcommand{\tick}{\checkmark}

\newcommand{\proofcase}[1]{\medskip\par\noindent\textbf{Case} #1:}

\renewcommand{\vec}[1]{\mathbf{#1}}

\newcommand{\IGNORE}[1]{}

\begin{document}

%\sloppy

\title[Partial Model Checking using Networks of LTS and Boolean Equation Systems]{Partial Model Checking using Networks of Labelled Transition Systems and Boolean Equation Systems}

\author[F.~Lang]{Fr\'ed\'eric Lang}
\address{{\sc Convecs} team, \INRIA\ {\em Grenoble -- Rh\^one-Alpes\/} and {\sc Lig} ({\em Laboratoire d'Informatique de Grenoble\/}), Montbonnot, France}
\email{\{Frederic.Lang,Radu.Mateescu\}@inria.fr}

\author[R.~Mateescu]{Radu Mateescu}
\address{\vspace{-18 pt}}
%\email{Radu.Mateescu@inria.fr}

%% mandatory lists of keywords and classifications:

% -------------------------------------------------------------------------- %

\begin{abstract}
  Partial model checking was proposed by Andersen in 1995 to verify a
  temporal logic formula compositionally on a composition of
  processes.  It consists in incrementally incorporating into the
  formula the behavioural information taken from one process --- an
  operation called quotienting --- to obtain a new formula that can be
  verified on a smaller composition from which the incorporated
  process has been removed.  Simplifications of the formula must be
  applied at each step, so as to maintain the formula at a tractable
  size.  In this paper, we revisit partial model checking.  First, we
  extend quotienting to the network of labelled transition systems
  model, which subsumes most parallel composition operators, including
  $m$-among-$n$ synchronisation and parallel composition using
  synchronisation interfaces, available in the \ELOTOS\ standard.
  Second, we reformulate quotienting in terms of a simple synchronous
  product between a graph representation of the formula (called
  formula graph) and a process, thus enabling quotienting to be
  implemented efficiently and easily, by reusing existing tools
  dedicated to graph compositions.  Third, we propose simplifications
  of the formula as a combination of bisimulations and reductions
  using Boolean equation systems applied directly to the formula
  graph, thus enabling formula simplifications also to be implemented
  efficiently.  Finally, we describe an implementation in the
  \CADP\ ({\em Construction and Analysis of Distributed Processes\/})
  toolbox and present some experimental results in which partial model
  checking uses hundreds of times less memory than on-the-fly model
  checking.
\end{abstract}

% -------------------------------------------------------------------------- %

\maketitle

% -------------------------------------------------------------------------- %
\section{Introduction}
\label{sec:intro}

	Concurrent safety critical systems can be verified using {\em model checking}~\cite{Clarke-Grumberg-Peled-00}, i.e., automatic evaluation of a temporal property against a formal model of the system.
	Although successful in many applications, model checking may face state explosion, particularly when the number of concurrent processes grows.

	State explosion can be tackled by {\em divide-and-conquer\/} approaches regrouped under the name {\em compositional verification}, which take advantage of the compositional structure of the concurrent system under verification.
	One such approach, which we call {\em compositional model generation\/} in this paper, consists in building the model of the system --- usually an \LTS\ ({\em Labelled Transition System\/}) --- in a stepwise manner, by successive compositions and minimisations modulo equivalence relations, possibly using {\em interface constraints\/}~\cite{Graf-Steffen-90,Krimm-Mounier-97} to avoid explosion of intermediate compositions.
	Tools using this approach~\cite{Garavel-Lang-01,Lang-05,Lang-06,Crouzen-Lang-11} are available in the \CADP\ ({\em Construction and Analysis of Distributed Processes\/})~\cite{Garavel-Lang-Mateescu-Serwe-11,Garavel-Lang-Mateescu-Serwe-12} toolbox.

        In this paper, we explore a dual approach named {\em partial model checking}, proposed by Andersen~\cite{Andersen-95,Andersen-Lind-Nielsen-99} for concurrent processes running asynchronously and composed using \CCS\ parallel composition and restriction operators.
	For a modal $\mu$-calculus~\cite{Kozen-83} formula $\varphi$ and a process composition $P_1 || \ldots || P_n$, Andersen uses an operation $\varphi/\!\!/P_1$ called {\em quotienting\/} of the formula $\varphi$ w.r.t. the process $P_1$, so that $P_1 || \ldots || P_n$ satisfies $\varphi$ if and only if the smaller composition $P_2 || \ldots || P_n$ satisfies $\varphi/\!\!/P_1$.
	In addition, simplifications can (and must) be applied to $\varphi/\!\!/P_1$ to reduce its size.
	Partial model checking is the incremental application of quotienting and simplifications, so that state explosion is avoided if the size of intermediate formulas can be kept sufficiently small.

	Partial model checking has been adapted and used successfully in various contexts, such as state-based models~\cite{Andersen-Staunstrup-Maretti-97a,Andersen-Staunstrup-Maretti-97b}, synchronous state/event systems~\cite{Bodentien-Vestergaard-Friis-Kristoffersen-Larsen-99}, and timed systems~\cite{Berard-Laroussinie-03,Cassez-Laroussinie-00,Laroussinie-Larsen-95,Laroussinie-Larsen-98,Larsen-Pettersson-Yi-95}.
	It has also been specialised for security properties~\cite{Martinelli-03}.
	More recently, it has been generalised to the full \CCS\ process algebra, with an application to the verification of parameterised systems~\cite{Basu-Ramakrishnan-03}.
	These various developments of partial model checking, although successful, were relatively scarce, which may be explained by the complexity of the method: obtaining a fully operational partial model checker requires a significant implementation effort and extensive experiments for fine-tuning and optimization.

	In this paper, we focus on partial model checking of the modal $\mu$-calculus applied to (untimed) concurrent asynchronous processes.
	By considering only binary associative parallel composition operators (such as \CCS\ and \CSP\ parallel compositions), previous works~\cite{Andersen-95,Andersen-Lind-Nielsen-99,Basu-Ramakrishnan-03} are not directly applicable to more general operators, such as $m$-among-$n$ synchronisation (where among $n$ processes executing in parallel, any $m$ of them must synchronise on a given action) and parallel composition by synchronisation interfaces (where all processes containing a given action in their synchronisation interface must synchronise on that action)~\cite{Garavel-Sighireanu-99}, present in the \ELOTOS\ standard and variants~\cite{Champelovier-Clerc-Garavel-et-al-11,ISO-15437}.
	Our first contribution in this paper is thus a generalisation of partial model checking to networks of {\LTS}s~\cite{Lang-05}, a general model that subsumes parallel composition, hiding, cutting, and renaming operators of standard process languages (\CCS, \CSP, \mCRL, \LOTOS, \ELOTOS, etc.), including the above-mentioned parallel composition operators.
	Regarding the communication of data values, our approach is applicable to classical (i.e., with static communication) value-passing process algebras equipped with early operational semantics.
	This framework encompasses a significant fragment of the $\pi$-calculus (containing channel mobility and bounded process creation), which can be translated into classical value-passing process algebras~\cite{Mateescu-Salaun-10}.

	In realistic cases, partial model checking handles huge formulas and processes, thus requiring efficient implementations.
	Our second contribution is a reformulation of quotienting as a synchronous product (which can itself be represented in the network model) between a graph representing the formula (called a {\em formula graph\/}) and the behaviour graph of a process, thus enabling efficient implementation using existing tools dedicated to graph manipulations.
	We prove that this reformulation is sound.
	Our third contribution is the reformulation of formula simplifications as a combination of graph reductions (including minimisations modulo equivalence relations and bisimulations) and partial evaluation of the formula graph using a \BES\ ({\em Boolean Equation System\/})~\cite{Andersen-94}.

	Verifying modal $\mu$-calculus formulas of arbitrary alternation depth is generally exponential in the size of the process graph, while verifying the alternation-free fragment remains of linear complexity.
	Our fourth contribution is a specialisation of the technique to alternation-free $\mu$-calculus formulas.
        We also present how this specialisation can be again generalised to handle also useful fairness operators of alternation 2 in linear time without developing the complex machinery to evaluate general alternation-2 $\mu$-calculus formulas.
	Finally, we present an implementation in \CADP\ and a case-study that illustrates the complementarity between partial and on-the-fly model checking.

\smallskip
\noindent
{\em Paper Overview.\/}
The modal $\mu$-calculus is presented in Section~\ref{sec:mu-calculus}.
The network of {\LTS}s model is presented in Section~\ref{sec:networks}.
The generalisation of quotienting to networks and its reformulation as a synchronous product is presented in Section~\ref{sec:reformulation}.
The simplification rules are presented in Section~\ref{sec:simplifications}.
The rules specific to alternation-free $\mu$-calculus formulas are presented in Section~\ref{sec:alternation-free}.
The way we handle fairness operators is presented in Section~\ref{sec:fairness}.
Our implementation of partial model checking of the regular alternation-free $\mu$-calculus extended with fairness operators is presented in Section~\ref{sec:implementation}.
Experimental results are presented in Section~\ref{sec:experimentation}.
Concluding remarks are given in Section~\ref{sec:conclusion}.
	This paper is an extended version of an earlier paper~\cite{Lang-Mateescu-12}.
%	Correctness proofs have been added in Section~\ref{sec:reformulation}, and Section~\ref{sec:fairness} is entirely new.

% -------------------------------------------------------------------------- %

\section{The Modal $\mu$-Calculus}
\label{sec:mu-calculus}

	We consider systems whose behavioural semantics can be represented using an \LTS\ ({\em Labelled Transition System\/}), and whose properties can be expressed in the modal $\mu$-calculus~\cite{Kozen-83}.

\begin{defi}[\LTS]
	An \LTS\ is a tuple $(\Sigma, A, \trans{}, s_0)$, where:

\begin{itemize}
	\item $\Sigma$ is a set of states,

	\item $A$ is a set of labels,

	\item $\trans{}\ \subseteq \Sigma \times A \times \Sigma$ is the (labelled) transition relation,

	\item and $s_0 \in \Sigma$ is the initial state.
\end{itemize}

	For an \LTS\ $S = (\Sigma, A, \trans{}, s_0)$, we may also write $s \trans{a} s' \in S$ (or simply $s \trans{a} s'$ when $S$ is clear from the context) instead of $(s, a, s') \in\ \rightarrow$.
\end{defi}

\begin{defi}[Syntax of the modal $\mu$-calculus]
	The modal $\mu$-calculus formulas ($\varphi$) are terms built from Boolean constants ($\false, \true$), Boolean connectors (disjunction $\lor$, conjunction $\land$, and negation $\neg$), modalities (possibility $\dia \_ \mond$ and necessity $\brac \_ \ket$), and fix-point operators (minimal $\mu$ and maximal $\nu$) over propositional variables $X$, generated by the following grammar:
\[
\begin{array}{rcl}
\varphi & ::= & \false ~\mid~ \varphi_1 \lor \varphi_2 ~\mid~ \dia a \mond \varphi_0~\mid~ \mu X . \varphi_0 \\
        & \mid & \true ~\mid~ \varphi_1 \land \varphi_2 ~\mid~ \brac a \ket \varphi_0 ~\mid~ \nu X . \varphi_0 \\
        & \mid & \neg \varphi_0 ~\mid~ X 
\end{array}
\]\bigskip

	\noindent To ensure a proper definition of fix-point operators, a commonly adopted and sufficient condition is that formulas $\varphi$ are {\em syntactically monotonic}~\cite{Kozen-83}, i.e., have an even number of negations on every path between a variable occurrence $X$ and the $\mu$ or $\nu$ operator that binds $X$.
        Therefore, we will only consider syntactically monotonic formulas.
        We write \Lmu\ for the set of $\mu$-calculus formulas.

	We write $\fv\,(\varphi)$ for the set of variables free in $\varphi$, and $\bv\,(\varphi)$ for  the set of variables bound in $\varphi$.
        We call a {\em closed formula\/} any formula $\varphi$ such that $\fv\,(\varphi) = \emptyset$.
	We assume that all bound variables have distinct names, and for $X \in \bv\,(\varphi)$, we write $\varphi[X]$ for the (unique) sub-formula of $\varphi$ of either form $\mu X.\varphi_0$ or $\nu X.\varphi_0$.
	Given $\varphi_1$ and $\varphi_2$, we write $\varphi_1[\varphi_2/X]$ for substituting all free occurrences of $X$ in $\varphi_1$ by $\varphi_2$ (while implicitly applying $\alpha$-conversion to maintain the unicity of bound variables).
\end{defi}

\begin{defi}[Semantics of the modal $\mu$-calculus]
	The semantics of the modal $\mu$-calculus are formally defined by the equations of Figure~\ref{fig:mc-semantics}.
	A propositional context $\rho$ is a partial function mapping propositional variables to sets of states and $\rho \oslash [ U / X ]$ stands for a propositional context identical to $\rho$ except that $X$ is mapped to $U$.
	The interpretation $\sem \varphi \antic \rho$ (also written $\sem \varphi \antic$ if $\rho$ is empty) of a state formula on an \LTS\ in a propositional context $\rho$ (which maps each variable free in $\varphi$ to a set of states) denotes the subset of states satisfying $\varphi$ in that context.
	The Boolean connectors are interpreted as usual in terms of set operations.
	The possibility modality $\dia a \mond \varphi_0$ (resp. the necessity modality $\brac a \ket \varphi_0$) denotes the states for which some (resp. all) of their outgoing transitions labelled by $a$ lead to states satisfying $\varphi_0$.
	The minimal fix-point operator $\mu X . \varphi_0$ (resp. the maximal fix-point operator $\nu X . \varphi_0$) denotes the least (resp. greatest) solution of the equation $X = \varphi_0$ interpreted over the complete lattice $\dia 2^\Sigma, \emptyset, \Sigma, \cap, \cup, \subseteq \mond$.
	A state $s$ satisfies a closed formula $\varphi$ if and only if $s \in \sem \varphi \antic$.
\end{defi}

\begin{figure}
\begin{center}
\begin{math}
\begin{array}{r@{\;}c@{\;}l}
\sem \false \antic \rho & = & \emptyset \\
	\sem \true \antic \rho & = & \Sigma \\
\sem \varphi_1 \lor \varphi_2 \antic \rho & = & \sem \varphi_1 \antic \rho \cup \sem \varphi_2 \antic \rho \\
	\sem \varphi_1 \land \varphi_2 \antic \rho & = & \sem \varphi_1 \antic \rho \cap \sem \varphi_2 \antic \rho \\
\sem \dia a \mond \varphi_0 \antic \rho & = & \{ s \in \Sigma \mid (\exists s'\in \Sigma)\ s \trans{a} s' \land s' \in \sem \varphi_0 \antic \rho \} \\
	\sem \brac a \ket \varphi_0 \antic \rho & = & \{ s \in \Sigma \mid (\forall s'\in \Sigma)\ s \trans{a} s' \implies s' \in \sem \varphi_0 \antic \rho \} \\
\sem \mu X . \varphi_0 \antic \rho & = & \bigcap \{ U \subseteq \Sigma \mid \sem \varphi_0 \antic (\rho \oslash [ U / X ]) \subseteq U \} \\
	\sem \nu X . \varphi_0 \antic \rho & = & \bigcup \{ U \subseteq \Sigma \mid U \subseteq \sem \varphi_0 \antic (\rho \oslash [ U / X ]) \} \\
\sem \neg \varphi_0 \antic \rho & = & \Sigma \setminus \sem \varphi_0 \antic \rho \\
\sem X \antic \rho & = & \rho (X) \\
\end{array}
\end{math}
\end{center}
\caption{Semantics of the modal $\mu$-calculus}
\label{fig:mc-semantics}
\end{figure}

\begin{prop}\label{prop:duality}
	The modal $\mu$-calculus satisfies the following identities:
\[
\begin{array}{rcl}
	\neg\true & = & \false \\
        \neg\false & = & \true \\
	\neg(\varphi_1 \land \varphi_2) & = & \neg \varphi_1 \lor \neg \varphi_2 \\
	\neg(\varphi_1 \lor \varphi_2) & = & \neg \varphi_1 \land \neg \varphi_2 \\
        \neg\brac a \ket \varphi_0 & = & \dia a \mond \neg \varphi_0 \\
        \neg\dia a \mond \varphi_0 & = & \brac a \ket \neg \varphi_0 \\
	\neg\nu X . \varphi_0 & = & \mu X . \neg \varphi_0 [ \neg X / X ] \\
	\neg\mu X . \varphi_0 & = & \nu X . \neg \varphi_0 [ \neg X / X ] \\
\end{array}
\]
\end{prop}

\begin{defi}[Positive form and disjunctive form]
	Every modal $\mu$-calculus formula $\varphi$ can be rewritten in both of the following forms:

\begin{itemize}
	\item A formula is in {\em positive form} if it contains any of the modal $\mu$-calculus operators but the negation operator $\neg$.
        Note that syntactic monotonicity implies that every negation can be eliminated using the identities of Proposition~\ref{prop:duality}.
	Given a modal $\mu$-calculus formula $\varphi$, we write $\varphi^+$ the corresponding formula in positive form.

	\item A formula is in {\em disjunctive form} if it contains only the constant $\false$, disjunctions, possibility modalities, minimal fix-points, propositional variables and negations.
	Every formula can be put in disjunctive form using the identities of Proposition~\ref{prop:duality}.
	Note that a formula in disjunctive form is not necessarily a disjunctive formula due to the presence of negations.
\end{itemize}
\end{defi}

\begin{defi}
	A formula $\varphi$ is {\em alternation-free\/} if $\varphi^+$ does not contain any sub-formula of the form $\mu X.\varphi_1$ (resp. $\nu X.\varphi_1$) containing a sub-formula of the form $\nu Y.\varphi_2$ (resp. $\mu Y.\varphi_2$) such that $X \in \fv\,(\varphi_2)$.
	The {\em fix-point sign\/} of a variable $X$ in $\varphi$ is $\mu$ (resp. $\nu$) if $\varphi^+[X]$ has the form $\mu X . \varphi_0$ (resp. $\nu X . \varphi_0$).
	We write \Lmui\ for the set of alternation-free $\mu$-calculus formulas, and more generally \Lmun\ for the set of $\mu$-calculus formulas of alternation up to $n$ (for some $n$).
\end{defi}

\begin{defi}[Block-labelled formula]
\label{def:block-labelled-formula}
	In this paper, we consider {\em block-labelled\/} formulas $\varphi$ in disjunctive form, in which each propositional variable $X$ is labelled by a natural number $k$, called its {\em block number}.

	Intuitively, a block-labelling is {\em well-formed\/} if the $\mu$-calculus formula can be converted into an equivalent set of $\mu$-calculus equations partitioned into blocks, so that all variables having the same block number are defined in the same block and if $k < k'$ then the equations within block number $k$ occur before the equations within block number $k'$.
	The proof is beyond the scope of this paper.
	The well-formedness conditions are the following:
\begin{enumerate}
	\item All occurrences of a given variable $X$ are labelled by the same block number $k$.

	\item All variables sharing the same block number have the same fix-point sign.

	\item For all $X^k \in \bv\,(\varphi), Y^{k'} \in \fv(\varphi[X^k])$ it holds that $k' \leq k$.
\end{enumerate}

	By convention, we assume without loss of generality that the even block numbers are associated to variables of sign $\mu$ and odd block numbers are associated to variables of sign $\nu$.

	Initially, every unlabelled formula $\varphi$ in disjunctive form can be turned into the well-formed block-labelled formula $\bl\,(\varphi, \true, 0, [])$, where $\bl\,(\psi, b, k, \gamma)$ is defined as follows, $\gamma$ denoting a mapping from variables to block numbers:
\[
\begin{array}{rcl}
\bl\,(\false, b, k, \gamma) & = & \false \\
\bl\,(X, b, k, \gamma) & = & X^{\gamma\,(X)} \\
\bl\,(\neg\varphi_0, b, k, \gamma) & = & \neg\bl\,(\varphi_0, \neg b, k, \gamma) \\
\bl\,(\varphi_1 \lor \varphi_2, b, k, \gamma) & = & \bl\,(\varphi_1, b, k, \gamma) \lor \bl\,(\varphi_2, b, k, \gamma) \\
\bl\,(\dia a \mond \varphi_0, b, k, \gamma) & = & \dia a \mond \bl\,(\varphi_0, b, k, \gamma) \\
\bl\,(\mu X . \varphi_0, b, k, \gamma) & = &
\left\{
\begin{array}{ll}
\mu X^k . \bl\,(\varphi_0, \true, k, \gamma[X \mapsto k]) & \mbox{if } b = \true \\
\mu X^{k+1} . \bl\,(\varphi_0, \true, k+1, \gamma[X \mapsto k+1]) & \mbox{otherwise}
\end{array}
\right.
\end{array}
\]
	
\noindent We write $\blocks(\varphi)$ for the set of block numbers occurring in $\varphi$.
	A block-labelled formula $\varphi$ in disjunctive form is {\em alternation-free\/} if $k' = k$ for all $X^k \in \bv(\varphi), Y^{k'} \in \fv(\varphi[X^k])$.
\end{defi}

	A well-known result of the $\mu$-calculus is that the variables of an alternation-free formula can be partitioned into blocks that have no cyclic dependencies.
	Another way to state this result is that any unlabelled formula in disjunctive form is alternation-free if and only if it can be block-labelled so that it satisfies the definition of alternation-free block-labelled formula.

	In the remainder of this paper, we will consider block-labelled formulas in disjunctive form.
	At last, we consider the following notion of formula equivalence, which is a slight generalisation of syntactic equality to enclose also the semantic notions of renaming, commutativity, and idempotence.

\begin{defi}
	Let $f$ be a bijective function from the set of propositional variables to itself, called a renaming.
%	Let $\varphi$ and $\varphi'$ be formulas in disjunctive form.
	For formulas in disjunctive form, we define syntactic equality modulo commutativity, idempotence and $f$-renaming as the smallest relation, written $=_f$, such that if $\varphi_i =_f \varphi_i'\ (i \in 0..2)$ then:
	\begin{itemize}
		\item $\false =_f \false$, $\neg\varphi_0 =_f \neg\varphi_0'$, $\dia a\mond\varphi_0 =_f \dia a\mond\varphi_0'$, $\varphi_1 \lor \varphi_2 =_f \varphi_1' \lor \varphi_2'$, $X =_f f(X)$, and $\mu X.\varphi_0 =_f \mu f(X).\varphi_0'$ for each propositional variable $X$ ({\em syntactic equality modulo renaming}),

		\item $\varphi_1 \lor \varphi_2 =_f \varphi_2' \lor \varphi_1'$ ({\em commutativity\/}),

		\item $\varphi_0 \lor \varphi_0 =_f \varphi_0'$ and $\varphi_0 =_f \varphi_0' \lor \varphi_0'$ ({\em idempotence\/}).
	\end{itemize}
\end{defi}

% -------------------------------------------------------------------------- %

\section{Networks of LTSs}
\label{sec:networks}

{\em Networks of LTSs\/} (or {\em networks\/} for short) are inspired from the {\sc Mec}~\cite{Arnold-89} and {\sc Fc2}~\cite{Bouali-Ressouche-Roy-deSimone-96} synchronisation vectors and were introduced in~\cite{Lang-05} as an intermediate model to represent compositions of {\LTS}s using various operators.
% We first give a few background definitions before defining the model formally.

\begin{defi}[Vector and vector projection]
	We write $n..m$ for the set of integers ranging from $n$ to $m$, or the empty set if $n > m$.
	A {\em vector\/} $\vec{v}$ of size $n$ is a total function on $1..n$.
	For $i \in 1..n$, we write $\vec{v}[i]$ for $\vec{v}$ applied to $i$, denoting the element of $\vec{v}$ stored at index $i$.
	We write $(e_1, \dots, e_n)$ for the vector $\vec{v}$ of size $n$ such that $(\forall i \in 1..n)\ \vec{v}[i] = e_i$.
	In particular, $()$ denotes a vector of size 0.

	Given $n \geq 1$ and $i \in 1..n$, $\vec{v}_{\setminus i}$ denotes the projection of $\vec{v}$ on to the set of indices $1..n \setminus \{i\}$, defined as the vector of size $n-1$ such that $(\forall j \in 1..i-1)\ \vec{v}_{\setminus i}[j] = \vec{v}[j]$ and $(\forall j \in i..n-1)\ \vec{v}_{\setminus i}[j] = \vec{v}[j+1]$.
\end{defi}

\begin{defi}[Network of LTSs]
A {\em network of LTSs\/} $N$ of size $n$ is a pair $(\vec{S}, V)$, where $\vec{S}$ is a vector of {\LTS}s (called {\em individual LTSs\/}) of size $n$, and $V$ is a set of {\em synchronisation rules}.
Each synchronisation rule has the form $(\vec{t}, a)$ with $a$ a label and $\vec{t}$ a vector of size $n$, called the {\em synchronisation vector}, of labels and occurrences of a special symbol $\bullet$ distinct from any label.
	Let $\vec{S}[i] = (\Sigma_i, A_i, \trans{}_i, s_i^0)\ (i \in 1..n)$.
	$N$ can be associated to a (global) \LTS\ $\lts\,(N)$ which is the parallel composition of individual {\LTS}s.
Each $(\vec{t}, a) \in V$ defines transitions labelled by $a$, obtained either by synchronisation (if more than one index $i$ is such that $\vec{t}[i] \neq \bullet$) or by interleaving (otherwise) of individual \LTS\ transitions.
Formally, $\lts\,(N) = (\Sigma, A, \trans{}, \vec{s}_0)$, where:

\begin{itemize}
	\item $\Sigma = \Sigma_1 \times \ldots \times \Sigma_n$,

	\item $A = \{ a \mid (\vec{t}, a) \in V \}$,

	\item $\vec{s}_0 = (s_1^0, \dots, s_n^0)$, and 

	\item $\trans{}$ is the relation satisfying $\vec{s} \trans{a} \vec{s}'$ if and only if there exists $(\vec{t}, a) \in V$ such that for all $i \in 1..n$:
\[
\left\{
\begin{array}{ll}
\vec{s}'[i] = \vec{s}[i] & \textup{if}\ \vec{t}[i] = \bullet \\
\vec{s}[i] \trans{\vec{t}[i]}_i \vec{s}'[i] & \textup{otherwise}
\end{array}
\right.
\]
\end{itemize}

	\noindent We write $A(\vec{t})$ for the set of {\em active\/} \LTS\ (indices), defined by $\{ i \mid i \in 1..n \land \vec{t}[i] \neq \bullet \}$.
\end{defi}

\begin{exa}\label{ex:network}
Let $a$, $b$, $c$, and $d$ be labels, and $P_1$, $P_2$, and $P_3$ be the processes defined in Figure~\ref{fig:global-lts} (top), where the initial states are denoted by bold circles.
Let $N = ((P_1, P_2, P_3), V)$ with $V =
\{ 	((a, a, \bullet), a),
	((a, \bullet, a), a),
	((b, b, b), b),$
	$((c, c, \bullet), \tau),
	((\bullet, \bullet, d), d) \}$,
whose global \LTS\ is depicted in Figure~\ref{fig:global-lts} (bottom left).
The first two rules express a nondeterministic synchronisation on $a$ between either $P_1$ and $P_2$, or $P_1$ and $P_3$.
The third rule expresses a multiway synchronisation on $b$.
The fourth rule yields an internal ($\tau$) transition.
The fifth rule expresses full interleaving of transitions labelled by $d$.
\end{exa}

\begin{figure}[t]
\begin{center}
\begin{tabular}{ccc}
\scalebox{0.75}{\input{p1.pstex_t}} &
\scalebox{0.75}{\input{p2.pstex_t}} &
\scalebox{0.75}{\input{p3.pstex_t}} \\
$P_1$ & $P_2$ & $P_3$
\end{tabular}

\vspace{5mm}
\begin{tabular}{cc}
\scalebox{0.75}{\input{compo.pstex_t}}
&
\scalebox{0.75}{\input{subcompo.pstex_t}} \\
$\lts\,(N)$ & $\lts\,(N_{\setminus 3})$ \\
\end{tabular}
\end{center}
\caption{Labelled Transition Systems for $N$ defined in Example~\ref{ex:network}}
\label{fig:global-lts}
\end{figure}

	The network of {\LTS}s model is used in the tool \EXPOPEN~\cite{Lang-05} of \CADP\ as an intermediate model for representing {\LTS}s composed using the hiding, renaming, cutting, and parallel composition operators present in the process algebras \CCS, \CSP, \LOTOS, and \mCRL, but also more expressive operators, such as $m$-among-$n$ synchronisation and parallel composition using synchronisation interfaces~\cite{Garavel-Sighireanu-99} present in \ELOTOS~\cite{ISO-15437} and \LOTOSNT~\cite{Champelovier-Clerc-Garavel-et-al-11}.
	For instance, the rules $\{((a, a, \bullet), a), ((a, \bullet, a), a), ((\bullet, a, a), a) \}$ realize 2-among-3 synchronisation on $a$.

	Computing the interactions of a process $P_i$ with its environment in a composition of processes $||_{j \in 1..n} P_j$ is easy when $||$ is a binary and associative parallel composition operator, since $||_{j \in 1..n} P_j = P_i\, ||\, (||_{j \in 1..n\setminus\{i\}} P_j)$.
	However, as argued in~\cite{Garavel-Sighireanu-99}, binary and associative parallel composition operators are of limited use when considering, e.g., $m$-among-$n$ synchronisation.
	A more involved operation named {\em sub-network extraction\/} is necessary for networks.

\begin{defi}[Sub-network extraction]
	$N = (\vec{S}, V)$ being a network of size $n$, we assume a function $\alpha\,(\vec{t}, a)$ that assigns a unique unused label to each $(\vec{t}, a) \in V$.
	Given $i \in 1..n$, we define $N_{\setminus i} = (\vec{S}_{\setminus i}, V_{\setminus i})$ the sub-network of $N$ modeling the environment of $\vec{S}[i]$ in $N$, where $V_{\setminus i} = \{ (\vec{t}_{\setminus i}, a) \mid (\vec{t}, a) \in V \land i \notin A(\vec{t}) \} \cup \{ (\vec{t}_{\setminus i}, \alpha\,(\vec{t}, a)) \mid (\vec{t}, a) \in V \land \{i\} \subset A(\vec{t}) \}$.
	$N$ is semantically equivalent to the network $((\vec{S}[i], \lts\,(N_{\setminus i})), V')$ with $V'$ the following set of rules, which define the interactions between $\vec{S}[i]$ and $N_{\setminus i}$:
\[
	\begin{array}{llllll}
	\{ & ((\bullet, & a), & a) & \mid (\vec{t}, a) \in V \land i \notin A(\vec{t}) & \} ~\cup \\
	\{ & ((\vec{t}[i], & \alpha\,(\vec{t}, a)), & a) & \mid (\vec{t}, a) \in V \land \{i\} \subset A(\vec{t}) & \} ~\cup \\
	\{ & ((a, & \bullet), & a) & \mid (\vec{t}, a) \in V \land \{i\} = A(\vec{t}) & \}
	\end{array}
\]
\end{defi}

	\noindent Each $\alpha(\vec{t}, a)$ is a unique interaction label between $\vec{S}[i]$ and $N_{\setminus i}$, which aims at avoiding erroneous interactions in case of nondeterministic synchronisation.

\begin{exa}
\label{ex:subnetwork}
	$N$ being defined in Example~\ref{ex:network}, $N_{\setminus 3}$ has vector of {\LTS}s $(P_1, P_2)$, $P_1$ and $P_2$ being defined in Figure~\ref{fig:global-lts} (top left and top middle), and rules \[\{((a, a), a), ((a, \bullet), \alpha_a), ((b, b), \alpha_b), ((c, c), \tau) \}\] with $\alpha_a = \alpha\,((a, \bullet, a), a)$ and $\alpha_b = \alpha\,((b, b, b), b)$;
	$\lts(N_{\setminus 3})$ is depicted in Figure~\ref{fig:global-lts} (bottom right);
	Composing it with $P_3$ using $\{ ((\bullet, a), a), ((a, \alpha_a), a), ((b, \alpha_b), b), ((\bullet, \tau), \tau), ((d, \bullet), d)\}$ yields $\lts(N)$.

	Note that if $a$ had been used instead of $\alpha_a$ in the above synchronisation rules, then the composition of $N_{\setminus 3}$ with $P_3$ would have enabled, in addition to the (correct) binary synchronisations on $a$ between $P_1$ and $P_2$ and between $P_1$ and $P_3$, the (incorrect) multiway synchronisation on $a$ between the three of $P_1, P_2$, and $P_3$.
	Indeed, the label $a$ resulting from the synchronisation between $P_1$ and $P_2$ in $N_{\setminus 3}$ --- rule $((a, a), a)$ in $N_{\setminus 3}$ --- could synchronise with the label $a$ in $P_3$ --- rule $((a, a), a)$ in the composition between $N_{\setminus 3}$ and $P_3$.
	Note however that $\vec{t}[i]$ can be used instead of $\alpha(\vec{t}, a)$ when the network does not have nondeterministic synchronisation on $\vec{t}[i]$, as is the case for $b$ and $\alpha_b$ in this example.
	In this paper we use $\alpha(\vec{t}, a)$ uniformly to avoid complications.
\end{exa}

% -------------------------------------------------------------------------- %

\section{Quotienting for Networks using Networks}
\label{sec:reformulation}

	To check a closed formula $\varphi$ on a network $N = (\vec{S}, V)$, one can choose an individual \LTS\ $\vec{S}[i]$, compute the quotient of the formula $\varphi$ with respect to $\vec{S}[i]$, and check the resulting quotient formula on the smaller (at least in number of individual {\LTS}s, but also hopefully in global \LTS\ size) network $N_{\setminus i}$.

\begin{defi}[Quotient formula]
	The quotient formula is written $\trad{\varphi}{i}{\emptyset}\ s^i_0$ and defined as follows for closed formulas in disjunctive form:
\[
\begin{array}{rcl}
\trad{\false}{i}{B}\ s & = & \false \\[2mm]
\trad{X^k}{i}{B}\ s & = & \trad{\varphi[X^k]}{i}{B}\ s \\[2mm]
\trad{(\notop \varphi_0)}{i}{B}\ s & = &
	\notop (\trad{\varphi_0}{i}{B}\ s) \\[2mm]
\trad{(\varphi_1 \lor \varphi_2)}{i}{B}\ s & = &
	(\trad{\varphi_1}{i}{B}\ s) \lor (\trad{\varphi_2}{i}{B}\ s) \\[2mm]
\trad{(\mu X^k. \varphi_0)}{i}{B}\ s & = &
\left \{
\begin{array}{l@{\quad}l}
X_s^k & \mbox{if } X_s^k \in B \\
\mu X_s^k . (\trad{\varphi_0}{i}{B \cup \{X_s^k\}}\ s) & \mbox{otherwise}
\end{array}
\right. \\[6mm]
\trad{(\langle a \rangle \varphi_0)}{i}{B}\ s & = &
	\bigvee_{(\vec{t}, a) \in V}\ 
	\left(
	\begin{array}{r@{}rlrl@{}ll}
	(& i \notin A(\vec{t}) & \land & \langle a \rangle & (\trad{\varphi_0}{i}{B}\ s) & ) & ~\lor \\
	(& \{i\} \subset A(\vec{t}) & \land \bigvee_{s \stackrel{\vec{t}[i]}{\longrightarrow_i} s'} & \langle \alpha\,(\vec{t}, a) \rangle & (\trad{\varphi_0}{i}{B}\ s') & ) & ~\lor \\
	(& \{i\} = A (\vec{t}) & \land \bigvee_{s \stackrel{\vec{t}[i]}{\longrightarrow_i} s'} & & (\trad{\varphi_0}{i}{B}\ s') & ) &
	\end{array}
	\right)
\end{array}
\]
\end{defi}\bigskip
% \caption{Quotienting for formulas in disjunctive form}
% \label{fig:quotienting}
% \end{figure}

	\noindent This definition follows and generalises Andersen's~\cite{Andersen-95} (specialised for \CCS) to networks.
	The main difference is the definition of $\trad{(\langle a \rangle \varphi_0)}{i}{B}\ s$, \CCS\ composition corresponding to vectors $((a, \bullet), a)$, $((\bullet, a), a)$, or $((a, \overline{a}), \tau)$, $a$ and $\overline{a}$ being an action and its \CCS\ {\em co-action}, making the use of special labels $\alpha(\vec{t}, a)$ not necessary.
	A minor difference is that we use $\mu$-calculus terms instead of equations\footnote{Note that terms will be compiled into graphs, thus enabling the sharing of sub-formulas that is also possible using equations.}.
	Any sub-formula produced by quotienting has the same block number as the original sub-formula, reflecting the order of equation blocks in Andersen's definition.
	The set $B$ keeps track of new variables already introduced in the quotient formula.
	Quotienting is well-defined, because formulas are finite, every $\varphi[X^k]$ has the form $\mu X^k.\varphi_0$ (because the formula is in disjunctive form), and the size of the set $B$ is bounded by $|\bv\,(\varphi)| \times |\Sigma_i|$.
	Note that well-formedness of the block-labelling is preserved by quotienting, because for every variable $X_s^k \in \bv\,(\trad{\varphi}{i}{\emptyset}\ s_0)$ we have $X^k \in \bv\,(\varphi)$ and for every variable $Y_{s'}^{k'} \in \fv\,((\trad{\varphi}{i}{\emptyset}\ s_0)[X_s^k])$ we have $Y^{k'} \in \fv\,(\varphi[X^k])$, and therefore $k' \leq k$.

\begin{exa}
\label{ex:quotienting}
	The $\mu$-calculus formula $\mu X^0. \langle a\rangle\true \lor \langle b\rangle X^0$ (existence of a path of zero or more $b$ leading to an $a$) can be rewritten to disjunctive form as $\mu X^0. \langle a\rangle\neg\false \lor \langle b\rangle X^0$.
	Quotienting of this formula with respect to $P_3$ in the network $N$ introduced in Example~\ref{ex:network} (page~\pageref{ex:network}) yields the formula $\mu X_0^0.\langle a\rangle\neg\false \lor \langle\alpha_a\rangle\neg\false \lor \langle\alpha_b\rangle\mu X_2^0.\langle a\rangle\neg\false \lor \false$.
	In other words, an action $a$ can be reached after a (possibly empty) sequence of $b$ actions in the network $N$ if and only if an action $a$, or an action $\alpha_a$, or an action $\alpha_b$ followed by an action $a$, can be reached immediately in $N_{\setminus 3}$, given the behaviour of $P_3$ depicted in Figure~\ref{fig:global-lts} (page~\pageref{fig:global-lts}).
\end{exa}

	We now show that quotienting can be implemented as a network that realises a product between an \LTS\ encoding the formula (called a {\em formula graph\/}) and an individual \LTS\ of the network under verification.

\begin{defi}[Circuit]
	Let $S = (\Sigma, A, \to, s_0)$ be an \LTS\ and $T \subseteq~\to$ be a subset of its transitions.
	The {\em states\/} of $T$ are defined as the set ${\rm st}\,(T) = \{ s, s' \in \Sigma \mid (s, \sigma, s') \in T \}$.
	$T$ is a {\em circuit\/} of $S$ if for all $s, s' \in {\rm st}\,(T)$ there is a sequence of transitions belonging to $T$ from $s$ to $s'$.
	A state $s \in {\rm st}\,(T)$ is a {\em root\/} of the circuit $T$ if there is a sequence of transitions from $s_0$ to $s$ that does not traverse any transition of $T$.
\end{defi}

\begin{defi}[Formula graph]
\label{def:formula-graph}
	A {\em formula graph\/} is an \LTS\ $(\Sigma, A, \to, s_0)$ such that:
\begin{enumerate}
	\item Every label $\sigma \in A$ has either form $\lor$, $\neg$, $\dia a \mond$ (for some $a$ belonging to a fixed set of action names), or $\mu^k$ (for some $k \in \mathbb{N}$).

	\item If $s_0 \trans{\delta} s \trans{\mu^k} s'$ for some $\delta \in A^*$ and $k \in \mathbb{N}$, then $k$ is even if and only if $\delta$ contains an even number of occurrences of the label $\neg$.

	\item If $s \in \Sigma$ is a root of a circuit then (a) the circuit contains a $\mu^{k}$-transition and (b) if the first $\mu^{k}$-transition traversed on the circuit starting in $s$ has block number $k'$ then every $\mu^k$-transition belonging to the circuit satisfies $k \geq k'$.
\end{enumerate}
\end{defi}

	\noindent Every formula graph can be decoded into a closed formula as follows.

\begin{defi}[Decoding a formula graph]
	A formula graph $P = (\Sigma, A, \rightarrow, s_0)$ encodes the modal $\mu$-calculus formula $\decs\,(P, s_0, \emptyset)$, where $\decs\,(P, s, E)$ is defined as follows ($E \subseteq \Sigma$).
	In our decoding every variable is uniquely identified by the source state $s$ and the block number $k$ of the $\mu$-transition, which we write $\holder{s}{k}$.
\[
\decs\,(P, s, E) = \bigvee_{s \trans{\sigma} s' \in P}\dect\,(P, s \trans{\sigma} s', E)
\]
\noindent where
\[
\begin{array}{rcl@{\quad}rcl}
\dect\,(P, s \trans{\lor} s', E) & = & \decs\,(P, s', E) \\[1mm]
\dect\,(P, s \trans{\neg} s', E) & = & \notop \decs\,(P, s', E) \\[1mm]
\dect\,(P, s \trans{\langle a \rangle} s', E) & = & \langle a \rangle \decs\,(P, s', E) \\[1mm]
\dect\,(P, s \trans{\mu^k} s', E) & = &
\left\{
\begin{array}{l@{\quad}l}
\holder{s}{k} & \mbox{if}\ s \in E \\
\mu \holder{s}{k}. \decs\,(P, s', E \cup \{s\}) & \mbox{otherwise}
\end{array}
\right.
\end{array}
\]
\end{defi}

	This definition implies that a deadlock state decodes as $\false$ (empty disjunction).
	Function $\decs$ is well-defined.
	In particular, it terminates because every cyclic path contains a label of the form $\mu^k$. By recording in the set $E$ the source states of traversed $\mu^k$-transitions, we thus avoid infinite traversals of cycles.
	In practice (see next section), formula graphs need not be decoded except for correctness proofs.

\begin{defi}[Encoding a formula into a formula graph]
	The formula graph corresponding to a formula $\varphi$ in disjunctive form is an \LTS\ written $\enc\,(\varphi)$, whose states are identified with sub-formulas of $\varphi$.
	The initial state of the formula graph is $\varphi$, $\false$ is a deadlock state, and each sub-formula has transitions as follows:
\[
\begin{array}{c@{\qquad}c@{\qquad}c}
X^k \trans{\lor} \varphi[X^k] &
\neg\varphi_0 \trans{\neg} \varphi_0 &
\langle a\rangle\varphi_0 \trans{\langle a\rangle} \varphi_0 \\
\varphi_1 \lor \varphi_2 \trans{\lor} \varphi_1 &
\varphi_1 \lor \varphi_2 \trans{\lor} \varphi_2 &
\mu X^k.\varphi_0 \trans{\mu^k} \varphi_0
\end{array}
\]
\end{defi}\bigskip

	\noindent Although the states of a formula graph are identified by formulas, only the transition labels are required for decoding.
	In figures, states will be simply identified by numbers.

	Note that the formula graph obtained by encoding a formula satisfies the conditions given in Definition~\ref{def:formula-graph}.
	Condition~(2) is a direct consequence of the block-labelling convention stated in Definition~\ref{def:block-labelled-formula}.
	Condition~(3) comes from the fact that the roots of the circuits are the states associated to formulas of the form $\mu X^k.\psi$ such that $X^k$ occurs free in $\psi$.
	In particular, subcondition~(b) is a consequence on the third well-formedness condition given in Definition~\ref{def:block-labelled-formula}.

\begin{figure}[t]
\begin{center}
\begin{tabular}{cc}
\scalebox{.75}{\input{formula-graph.pstex_t}} &
\scalebox{.75}{\input{quotient-graph.pstex_t}} \\
(a) & (b) \\
\multicolumn{2}{c}{\scalebox{.75}{\input{quotient-graph-simplified.pstex_t}}} \\
\multicolumn{2}{c}{(c)}
\end{tabular}
\end{center}\vspace{-4mm}
\caption{Examples of formula graphs}
\label{fig:formula-graphs}
\end{figure}

\begin{exa}
\label{ex:formula-graph}
	The formula graph corresponding to the formula $\mu X^0. (\langle a\rangle\true) \lor \langle b\rangle X^0$ introduced in Example~\ref{ex:quotienting} is depicted in Figure~\ref{fig:formula-graphs} (a).
\end{exa}

	We now prove that our encoding of closed formulas into formula graphs is sound, in the sense that the formula can be recovered from the formula graph into which the formula is encoded.
        This is stated formally in Proposition~\ref{prop:encoding-soundness} below, which is a corollary of the following Lemma:

\begin{lem}\label{lem:encoding-soundness}
        Let $\varphi$ be a closed formula in disjunctive form and $f$ be a renaming that maps each propositional variable $X^k \in \bv\,(\varphi)$ to $\holder{\varphi[X^k]}{k}$.
	For every sub-formula $\psi$ of $\varphi$, if $\{ \varphi[Y^k] \mid Y^k \in \fv\,(\psi) \} \subseteq E$ and $E \cap \{ \varphi[Y^k] \mid Y^k \in \bv\,(\psi) \} = \emptyset$, then $\decs\,(\enc\,(\varphi), \psi, E) =_f \psi$.
\end{lem}

\proof
	We proceed by structural induction on $\psi$:

\proofcase{$\psi = \false$}
	By definition of $\enc\,(\varphi)$, the state $\psi$ has no outgoing transition.
	Therefore by definition of $\decs$, we have $\decs\,(\enc\,(\varphi), \false, E) = \false$.

\proofcase{$\psi = X^k$}
	By definition of $\enc\,(\varphi)$, the state $\psi$ has a single transition $X^k \trans{\lor} \varphi[X^k]$.
	Therefore by definition of $\decs$, we have $\decs\,(\enc\,(\varphi), X^k, E) = \decs\,(\enc\,(\varphi), \varphi[X^k], E)$.
	Since $X^k \in \fv\,(X^k)$, by the hypothesis $\varphi[X^k] \in E$.
	It follows by definition of $\decs$ that $\decs\,(\enc\,(\varphi), X^k, E) = \holder{\varphi[X^k]}{k} =_f X^k$.

\proofcase{$\psi = \neg\psi_0$}
	By definition of $\enc\,(\varphi)$, the state $\psi$ has a single transition $\neg\psi_0 \trans{\neg} \psi_0$.
	Therefore by definition of $\decs$, we have $\decs\,(\enc\,(\varphi), \neg\psi_0, E) = \neg\decs\,(\enc\,(\varphi), \psi_0, E)$.
	Since $\fv\,(\psi_0) = \fv\,(\psi)$ and $\bv(\psi_0) = \bv\,(\psi)$, the induction hypothesis holds and then $\decs\,(\enc\,(\varphi), \psi_0, E) =_f \psi_0$.
	It follows immediately that $\decs\,(\enc\,(\varphi), \neg\psi_0 , E) =_f \neg\psi_0$.

\proofcase{$\psi = \psi_1 \lor \psi_2$}
	By definition of $\enc\,(\varphi)$, the state $\psi$ has two transitions $\psi_1 \lor \psi_2 \trans{\lor} \psi_1$ and $\psi_1 \lor \psi_2 \trans{\lor} \psi_2$.
	Therefore by definition of $\decs$, we have $\decs\,(\enc\,(\varphi), \psi_1 \lor \psi_2, E) = \decs\,(\enc\,(\varphi), \psi_1, E) \lor \decs\,(\enc\,(\varphi), \psi_2, E)$ (modulo commutativity if the transitions are enumerated in the opposite order, and idempotence if the transitions are identical).
	Since $\fv\,(\psi_1) \cup \fv\,(\psi_2) = \fv\,(\psi)$ and $\bv(\psi_1) \cup \bv\,(\psi_2) = \bv\,(\psi)$, the induction hypothesis holds and then we have both $\decs\,(\enc\,(\varphi), \psi_1, E) =_f \psi_1$ and $\decs\,(\enc\,(\varphi), \psi_2, E) =_f \psi_2$.
	It follows that $\decs\,(\enc\,(\varphi), \psi_1 \lor \psi_2, E) =_f \psi_1 \lor \psi_2$.

\proofcase{$\psi = \langle a\rangle\psi_0$}
	By definition of $\enc\,(\varphi)$, the state $\psi$ has a single transition $\langle a\rangle\psi_0 \trans{\langle a\rangle} \psi_0$.
	Therefore by definition of $\decs$, we have $\decs\,(\enc\,(\varphi), \langle a\rangle\psi_0, E) = \langle a\rangle\decs\,(\enc\,(\varphi), \psi_0, E)$.
	Since $\fv\,(\psi_0) = \fv\,(\psi)$ and $\bv(\psi_0) = \bv\,(\psi)$, the induction hypothesis holds and then $\decs\,(\enc\,(\varphi), \psi_0, E) =_f \psi_0$.
	It follows immediately that $\decs\,(\enc\,(\varphi), \langle a\rangle\psi_0, E) =_f \langle a\rangle\psi_0$.

\proofcase{$\psi = \mu X^k.\psi_0$}
	By definition of $\enc\,(\varphi)$, the state $\psi$ has a single transition $\mu X^k.\psi_0 \trans{\mu^k} \psi_0$.
	Also, $\mu X^k.\psi_0 \notin E$ because $\mu X^k.\psi_0 = \varphi[X^k]$, $X^k \in \bv\,(\psi)$ and, by hypothesis, $E \cap \{\varphi[Y^k] \mid Y^k \in \bv\,(\psi)\} = \emptyset$.
	As a consequence and by definition of $\decs$, we have \[\decs\,(\enc\,(\varphi), \mu X^k.\psi_0, E) = \mu \holder{\mu X^k.\psi_0}{k}.\decs\,(\enc\,(\varphi), \psi_0, E \cup \{\mu X^k.\psi_0 \}).\]
	Since $\mu X^k.\psi_0 = \varphi[X^k]$, the latter formula is also equal to $\mu \holder{\varphi[X^k]}{k}.\decs\,(\enc\,(\varphi), \psi_0, E \cup \{\varphi[X^k] \})$.
	To apply the induction hypothesis, we must show that $\{\varphi[Y^k] \mid Y^k \in \fv\,(\psi_0)\} \subseteq E \cup \{\varphi[X^k] \}$ and that $(E \cup \{\varphi[X^k]\}) \cap \{\varphi[Y^k] \mid Y^k \in \bv\,(\psi_0)\} = \emptyset$.
	This is true by hypothesis and because $\fv\,(\psi_0) = \fv\,(\psi) \cup \{X^k\}$ and $\bv\,(\psi_0) = \bv\,(\psi) \setminus \{X^k\}$.
	Therefore, $\decs\,(\enc\,(\varphi), \psi_0, E) =_f \psi_0$.
	It follows immediately that $\decs\,(\enc\,(\varphi), \mu X^k.\psi_0, E) =_f \mu X^k.\psi_0$.
\qed

\begin{prop}
\label{prop:encoding-soundness}
	If $\varphi$ is a closed formula in disjunctive form, then
	$\decs\,(\enc\,(\varphi), \varphi, \emptyset) =_f \varphi$
	where $f$ maps each propositional variable $X^k \in \bv\,(\varphi)$ to $\holder{\varphi[X^k]}{k}$.
\end{prop}

\proof
	If $\varphi$ is a closed formula, then $\fv\,(\varphi) = \emptyset$.
	We have $\{ \varphi[Y^k] \mid Y^k \in \fv\,(\varphi) \} = \emptyset$.
	Therefore, the hypotheses of Lemma~\ref{lem:encoding-soundness} are satisfied, which implies $\decs\,(\enc\,(\varphi), \varphi, \emptyset) =_f \varphi$.
\qed

	Using this encoding, the quotient of a formula with respect to the $i$th \LTS\ of a network can be computed as a synchronous product using a network called {\em quotient formula network}.

\begin{defi}[Quotient formula network]
	Let $\varphi$ be a modal $\mu$-calculus formula in disjunctive form, $N = (\vec{S}, V)$ be a network of size $n$, and $i \in 1..n$.
	The {\em quotient formula network} of $\varphi$ with respect to $\vec{S}[i]$ is defined as the network $((\enc\,(\varphi), \vec{S}[i]), V\dblslash_i)$, where $V\dblslash_i$ denotes the following set of rules:
\[
\begin{array}{lllll}
\{~ ((\sigma, & \bullet), & \sigma) &
	\mid \sigma \in \{\neg, \lor\} \cup \{\mu^k \mid k \in \blocks(\varphi)\}~\} & ~\cup \\
\{~ ((\langle a\rangle, & \bullet), & \langle a\rangle) &
	\mid (\vec{t}, a) \in V \land i \notin A(\vec{t}) ~\} & ~\cup \\
\{~ ((\langle a\rangle, & \vec{t}[i]), & \langle\alpha\,(\vec{t}, a)\rangle) &
	\mid (\vec{t}, a) \in V \land \{i\} \subset A(\vec{t}) ~\} & ~\cup \\
\{~ ((\langle a\rangle, & \vec{t}[i]), & \lor) &
	\mid (\vec{t}, a) \in V \land \{i\} = A(\vec{t}) ~\}
\end{array}
\]
\end{defi}\bigskip

	\noindent Note that the \LTS\ corresponding to the quotient formula network is a formula graph.
	This can easily be shown by observing that, if $(\psi_1, s_1) \trans{\delta} (\psi_n, s_n)$ is a transition sequence of the quotient formula network, then there exists a transition sequence of the form $\psi_1 \trans{\delta'} \psi_n$ in the input formula graph, such that the $\mu$-projection of $\delta'$ (i.e., the sequence obtained from $\delta'$ by keeping only the $\mu^k$-labels) and the $\mu$-projection of $\delta$ are identical.
	In addition, if the transition sequence labelled by $\delta$ is a circuit, then $\delta'$ can be found such that the transition sequence labelled by $\delta'$ is also a circuit.
	This ensures that conditions~(2) and~(3) of Definition~\ref{def:formula-graph} are preserved in the \LTS\ corresponding to the quotient formula network.

	We now prove that the \LTS\ corresponding to the quotient formula network indeed encodes the quotient correctly.
        This is stated formally in Proposition~\ref{lem:encoding-soundness} below, which is a corollary of the following Lemma:

\begin{lem}\label{lem:correctness}
	Let $\varphi$ be a closed formula in disjunctive form, $N = (\vec{S}, V)$ be a network of size $n$, $i \in 1..n$, $P = \lts\,((\enc\,(\varphi), \vec{S}[i]), V\dblslash_i)$ be the quotient formula network of $\varphi$ with respect to $\vec{S}[i]$, $s$ be a state of $\vec{S}[i]$, and $f$ be a renaming that maps each propositional variable $Y_t^k \in \bv\,(\trad{\varphi}{i}{B}\ s_0^i)$ to $\holder{(\varphi[Y^k], t)}{k}$.
	If $E = \{ (\varphi[Y^k], t) \mid Y_t^k \in B\}$ then for every sub-formula $\psi$ of $\varphi$, $\decs\,(P, (\psi, s), E) =_f \trad{\psi}{i}{B}\ s$.
\end{lem}

\proof
	We proceed by case on $\psi$ and by structural induction on the formula $\trad{\psi}{i}{B}\ s$ (which is finite):

\proofcase{$\psi = \false$}
	By definition of $P$, the state $(\false, s)$ has no outgoing transition, because by definition of $\enc\,(\varphi)$ the state $\false$ has no outgoing transition, and $V\dblslash_i$ contains no synchronisation rule of the form $((\bullet, a), b)$.
	Therefore, by definition of $\decs$ we have $\decs\,(P, (\false, s), E) = \false$ and by definition of quotienting we have $\trad{\false}{i}{B}\ s = \false$.
	It follows immediately that $\decs\,(P, (\false, s), E) =_f \trad{\false}{i}{B}\ s$.

\proofcase{$\psi = X^k$}
	By definition of $P$, the state $(X^k, s)$ has a transition $(X^k, s) \trans{\lor} (\varphi[X^k], s)$, because by definition of $\enc\,(\varphi)$ the state $X^k$ has a transition $X^k \trans{\lor} \varphi[X^k]$ and $V\dblslash_i$ contains the synchronisation rule $((\lor, \bullet), \lor)$.
	The state $(X^k, s)$ has no other transition in $P$, because the state $X^k$ has no other transition and $V\dblslash_i$ does not contain other synchronisation rules of either form $((\bullet, a), b)$ or $((\lor, a), b)$.
	Therefore, we have $\decs\,(P, (X^k, s), E) = \decs\,(P, (\varphi[X^k], s), E)$ by definition of $\decs$.
	As formulas are in disjunctive form, $\varphi[X^k]$ has the form $\mu X^k.\psi_0$.
	The rest of the proof for this case is identical to the case $\psi = \mu X^k.\psi_0$ detailed below.

\proofcase{$\psi = \neg\psi_0$}
	By definition of $P$, the state $(\neg\psi_0, s)$ has a transition $(\neg\psi_0, s) \trans{\neg} (\psi_0, s)$, because by definition of $\enc\,(\varphi)$ the state $\neg\psi_0$ has a transition $\neg\psi_0 \trans{\neg} \psi_0$ and $V\dblslash_i$ contains the synchronisation rule $((\neg, \bullet), \neg)$.
	The state $(\neg\psi_0, s)$ has no other transition in $P$, because the state $\neg\psi_0$ has no other transition and $V\dblslash_i$ does not contain other synchronisation rules of either form $((\bullet, a), b)$ or $((\neg, a), b)$.
	On the one hand, we thus have $\decs\,(P, (\neg\psi_0, s), E) = \neg\decs\,(P, (\psi_0, s), E)$ by definition of $\decs$.
	On the other hand, we have $\trad{(\neg\psi_0)}{i}{B}\ s = \neg(\trad{\psi_0}{i}{B}\ s)$ by definition of quotienting.
	Also $\trad{\psi_0}{i}{B}\ s$ is a proper sub-formula of $\trad{\psi}{i}{B}\ s$.
	Therefore, by induction hypothesis we have $\decs\,(P, (\psi_0, s), E) =_f \trad{\psi_0}{i}{B}\ s$.
	It follows immediately that $\decs\,(P, (\neg\psi_0, s), E) =_f \trad{(\neg\psi_0)}{i}{B}\ s$.

\proofcase{$\psi = \psi_1 \lor \psi_2$}
	By definition of $P$, the state $(\psi_1 \lor \psi_2, s)$ has transitions $(\psi_1 \lor \psi_2, s) \trans{\lor} (\psi_1, s)$ and $(\psi_1 \lor \psi_2, s) \trans{\lor} (\psi_2, s)$, because by definition of $\enc\,(\varphi)$ the state $\psi_1 \lor \psi_2$ has transitions $\psi_1 \lor \psi_2 \trans{\lor} \psi_1$ and $\psi_1 \lor \psi_2 \trans{\lor} \psi_2$ and $V\dblslash_i$ contains the synchronisation rule $((\lor, \bullet), \lor)$.
	The state $(\psi_1 \lor \psi_2, s)$ has no other transition in $P$, because the state $\psi_1 \lor \psi_2$ has no other transition and $V\dblslash_i$ does not contain other synchronisation rules of either form $((\bullet, a), b)$ or $((\lor, a), b)$.
	On the one hand, we thus have $\decs\,(P, (\psi_1 \lor \psi_2, s), E) = \decs\,(P, (\psi_1, s), E) \lor \decs\,(P, (\psi_2, s), E)$ by definition of $\decs$.
	On the other hand, we have $\trad{(\psi_1 \lor \psi_2)}{i}{B}\ s = (\trad{\psi_1}{i}{B}\ s) \lor (\trad{\psi_2}{i}{B}\ s)$ by definition of quotienting.
	Also $\trad{\psi_1}{i}{B}\ s$ and $\trad{\psi_2}{i}{B}\ s$ are proper sub-formulas of $\trad{\psi}{i}{B}\ s$.
	Therefore, by induction hypothesis we have $\decs\,(P, (\psi_1, s), E) =_f \trad{\psi_1}{i}{B}\ s$ and $\decs\,(P, (\psi_2, s), E) =_f \trad{\psi_2}{i}{B}\ s$.
	It follows immediately that $\decs\,(P, (\psi_1 \lor \psi_2, s), E) =_f \trad{(\psi_1 \lor \psi_2)}{i}{B}\ s$.

\proofcase{$\psi = \langle a\rangle\psi_0$}
	By definition of $\enc\,(\varphi)$, the state $\langle a\rangle\psi_0$ has a transition $\langle a\rangle\psi_0 \trans{\langle a\rangle} \psi_0$.
	By definition of $P$, the state $(\langle a\rangle\psi_0, s)$ has three kinds of transitions:
	\begin{itemize}
		\item A transition of the form $(\langle a\rangle\psi_0, s) \trans{\langle a\rangle} (\psi_0, s)$ for each $(\vec{t}, a) \in V$ such that $i \notin A(\vec{t})$, because $V\dblslash_i$ contains the synchronisation rule $((\langle a\rangle, \bullet), \langle a\rangle)$.
		This corresponds to a disjunct of the form $i \notin A(\vec{t}) \land \langle a\rangle(\trad{\psi_0}{i}{B}\ s)$ in the definition of $\trad{(\langle a\rangle\psi_0)}{i}{B}\ s$.

		\item A transition of the form $(\langle a\rangle\psi_0, s) \trans{\langle \alpha\,(\vec{t}, a)\rangle} (\psi_0, s')$ for each $(\vec{t}, a) \in V$ such that $\{i\} \subset A(\vec{t})$ and for each transition $s \trans{\vec{t}[i]}_i s'$ in $\vec{S}[i]$, because $V\dblslash_i$ contains the synchronisation rule $((\langle a\rangle, \vec{t}[i]), \langle\alpha\,(\vec{t}, a)\rangle)$.
		This corresponds to a disjunct of the form $\{i\} \subset A(\vec{t}) \land \bigvee_{s \trans{\vec{t}[i]}_i s'} \langle\alpha\,(\vec{t}, a)\rangle(\trad{\psi_0}{i}{B}\ s')$ in the definition of $\trad{(\langle a\rangle\psi_0)}{i}{B}\ s$.

		\item A transition of the form $(\langle a\rangle\psi_0, s) \trans{\lor} (\psi_0, s')$ for each $(\vec{t}, a) \in V$ such that $\{i\} = A(\vec{t})$ and for each transition $s \trans{\vec{t}[i]}_i s'$ in $\vec{S}[i]$, because $V\dblslash_i$ contains the synchronisation rule $((\langle a\rangle, \vec{t}[i]), \lor)$.
		This corresponds to a disjunct of the form $\{i\} = A(\vec{t}) \land \bigvee_{s \trans{\vec{t}[i]}_i s'} (\trad{\psi_0}{i}{B}\ s')$ in the definition of $\trad{(\langle a\rangle\psi_0)}{i}{B}\ s$.
	\end{itemize}

	\noindent The state $(\langle a\rangle\psi_0, s)$ has no other transitions in $P$, because the state $\langle a\rangle\psi_0$ has no other transition and $V\dblslash_i$ does not contain other synchronisation rules of either form $((\bullet, b), c)$ or $((\langle a\rangle, b), c)$.
	Also, $\trad{\psi_0}{i}{B}\ s$ and $\trad{\psi_0}{i}{B}\ s'$ are proper sub-formulas of $\trad{\psi}{i}{B}\ s$.
	By induction hypothesis, we have $\decs\,(P, (\psi_0, s), E) =_f \trad{\psi_0}{i}{B}\ s$ and $\decs\,(P, (\psi_0, s'), E) =_f \trad{\psi_0}{i}{B}\ s'$.
	It then follows immediately that $\decs\,(P, (\langle a\rangle\psi_0, s), E) =_f \trad{(\langle a\rangle\psi_0)}{i}{B}\ s$.

%\proofcase{$\psi = \mu X^k.\psi_0$} By definition of $P$, the state
%$(\mu X^k.\psi_0, s)$ has a transition $(\mu X^k.\psi_0, s)
%\trans{\mu^k} (\psi_0, s)$, because by definition of $\enc\,(\varphi)$
%the state $\mu X^k.\psi_0$ has a transition $\mu X^k.\psi_0
%\trans{\mu^k} \psi_0$ and $V\dblslash_i$ contains the synchronisation
%rule $((\mu^k, \bullet), \mu^k)$.  The state $(\mu X^k.\psi_0, s)$ has
%no other transition in $P$, because the state $\mu X^k.\psi_0$ has no
%other transition and $V\dblslash_i$ does not contain other
%synchronisation rules of either form $((\bullet, a), b)$ or $((\mu,
%a), b)$.  We consider two cases:
\proofcase{$\psi = \mu X^k.\psi_0$} By definition of $P$, and since by
definition of $\enc\,(\varphi)$ the state $\mu X^k.\psi_0$ has a
transition $\mu X^k.\psi_0 \trans{\mu^k} \psi_0$ and $V\dblslash_i$
contains the synchronisation rule $((\mu^k, \bullet), \mu^k)$, the
state $(\mu X^k.\psi_0, s)$ has a transition $(\mu X^k.\psi_0, s)
\trans{\mu^k} (\psi_0, s)$.  The state $(\mu X^k.\psi_0, s)$ has no
other transition in $P$, because the state $\mu X^k.\psi_0$ has no
other transition and $V\dblslash_i$ does not contain other
synchronisation rules of either form $((\bullet, a), b)$ or $((\mu,
a), b)$.  We consider two cases:
	\begin{itemize}
		\item If $(\mu X^k.\psi_0, s) \in E$ then by hypothesis $X_s^k \in B$.
		On the one hand, we thus have $\decs\,(P, (\mu X^k.\psi_0, s), E) = \holder{(\mu X^k.\psi_0, s)}{k}$ by definition of $\decs$.
		On the other hand, we have $\trad{(\mu X^k.\psi_0)}{i}{B}\ s = X_s^k$ by definition of quotienting.
		We also have $\holder{(\mu X^k.\psi_0, s)}{k} =_f X^k$ by definition of $=_f$ and because $\mu X^k.\psi_0 = \varphi[X^k]$.
		It follows immediately that $\decs\,(P, (\mu X^k.\psi_0, s), E) =_f \trad{(\mu X^k.\psi_0)}{i}{B}\ s$.

		\item If  $(\mu X^k.\psi_0, s) \notin E$ then by hypothesis $X_s^k \notin B$.
		On the one hand, we thus have $\decs\,(P, (\mu X^k.\psi_0, s), E) = \mu \holder{(\mu X^k.\psi_0, s)}{k}.\decs\,(P, (\psi_0, s), E')$ where $E' = E \cup \{\mu X^k.\psi_0\}$, by definition of $\decs$.
		On the other hand, we have $\trad{(\mu X^k.\psi_0)}{i}{B}\ s = \mu X_s^k.(\trad{\psi_0}{i}{B \cup \{X_s^k\}}\ s)$ by definition of quotienting.
		Also, $\trad{\psi_0}{i}{B \cup \{X_s^k\}}\ s$ is a proper sub-formula of $\trad{\psi}{i}{B}\ s$.
		By induction hypothesis, we thus have $\decs\,(P, (\psi_0, s), E') =_f \trad{\psi_0}{i}{B \cup \{X_s^k\}}\ s$ using $E' = E \cup \{(\mu X^k.\psi_0, s)\} = \{(\varphi[Y^k], t) \mid Y_t^k \in B \cup \{X_s^k\}\}$.
		It then follows immediately that $\decs\,(P, (\mu X^k.\psi_0, s), E) =_f \trad{(\mu X^k.\psi_0)}{i}{B}\ s$.
	\end{itemize}
\qed

\begin{prop}\label{prop:correctness}
	The \LTS\ corresponding to the quotient formula network of $\varphi$ with respect to $\vec{S}[i]$ encodes the quotient of $\varphi$ with respect to $\vec{S}[i]$.
\end{prop}

\proof
	Let $P$ be the quotient formula network of $\varphi$ with respect to $\vec{S}[i]$, in other words, $P = \lts\,((\enc\,(\varphi), \vec{S}[i]), V\dblslash_i)$.
	Since $\{ (\varphi[Y^k], t) \mid Y_t^k \in \emptyset\} = \emptyset$, then we have by Lemma~\ref{lem:correctness} that $\decs\,(P, (\varphi, s_0^i), \emptyset) =_f \trad{\varphi}{i}{\emptyset}\ s_0^i$, where $f$ maps each propositional variable $Y_t^k \in \bv\,(\trad{\varphi}{i}{B}\ s_0^i)$ to $\holder{(\varphi[Y^k], t)}{k}$.
	In other words $P$, the quotient formula network of $\varphi$ with respect to $\vec{S}[i]$, encodes $\trad{\varphi}{i}{\emptyset}\ s_0^i$, which is the quotient of $\varphi$ with respect to $\vec{S}[i]$.
\qed

\begin{exa}
\label{ex:quotient-example}
	Consider the network $N$ of Example~\ref{ex:network} (page~\pageref{ex:network}) and the formula of Example~\ref{ex:formula-graph} (page~\pageref{ex:formula-graph}).
	Quotienting of the formula with respect to $P_3$ involves the following set of rules:\\
\centerline{$\{((\neg, \bullet), \neg), ((\lor, \bullet), \lor), ((\mu^0, \bullet), \mu^0), ((\langle a\rangle, \bullet), \langle a\rangle), ((\langle a\rangle, a), \langle \alpha_a\rangle), ((\langle b\rangle, b), \langle \alpha_b\rangle) \}$}
	It yields the formula graph depicted in Figure~\ref{fig:formula-graphs} (b), page~\pageref{fig:formula-graphs}.
	This graph encodes as expected the quotient formula of Example~\ref{ex:quotienting} (page~\pageref{ex:quotienting}), which can be evaluated on $N_{\setminus 3}$.
\end{exa}

	Working with formulas in disjunctive form is crucial: branches in the formula graph denote disjunctions between sub-formulas ({\em or-nodes\/}).
	During composition between the formula graph and an individual \LTS, the impossibility to synchronise on a modality $\langle a\rangle$ (no transition labelled by $\vec{t}[i]$ in the current state of the individual \LTS) denotes invalidation of the corresponding sub-formula, which merely disappears, in conformance with the equality $\false \lor \varphi_0 = \varphi_0$.

% -------------------------------------------------------------------------- %

\section{Formula Graph Simplifications}
\label{sec:simplifications}

	The quotient of a formula graph with $n$ states with respect to an \LTS\ with $m$ states may have up to $n \times m$ states.
	Hence, as observed by Andersen~\cite{Andersen-95}, simplifications are needed to keep intermediate quotiented formulas at a reasonable size.
	We present in Figure~\ref{tab:simplification-rules} several simplifications applying to formula graphs, as conditional rules of the form ``$l \leadsto r\ (\mathit{cond})$'' where $l$ and $r$ are transition relations and $\mathit{cond}$ is a Boolean condition.
	$l$, $r$, and $\mathit{cond}$ are expressed using variables representing either states (written $s, s_1, s_2, \ldots$) or labels (written $\sigma, \sigma_1, \sigma_2, \ldots$), such that every variable occurring in $r$ or in $\mathit{cond}$ must also occur in $l$.
	It means that all transitions matching the left-hand side so that $\mathit{cond}$ is satisfied can be replaced by the transitions of the right-hand side.

\begin{figure}[t]
\[
\begin{array}{l@{\quad}r@{\quad}c@{\quad}l@{\quad}l}
\hline
(1) &
	\xymatrix@C=.5pc@R=.8pc{
		& s_1 \ar[d]^\lor \\
		& s_2 \ar[dl]_{\sigma_3} \ar[dr]^{\sigma_n} & \\
		s_3 & \ldots & s_n
	}
	& \leadsto &
	\xymatrix@C=.5pc@R=.8pc{
		& s_1 \ar@/_1pc/[ddl]_{\sigma_3} \ar@/^1pc/[ddr]^{\sigma_n} & \\
		& s_2 \ar[dl]_{\sigma_3} \ar[dr]^{\sigma_n} & \\
		s_3 & \ldots & s_n
        }
        & \txt<5cm>{($s_3, \ldots, s_n$ are all the successors of $s_2$)} \\
\hline
(2) & \xymatrix{s_1 \ar@(dl, dr)[]_{\mu^k}} 
	& \leadsto & s_1 \\
\hline
(3) & \xymatrix{s_1 \ar[r]^\neg & s_2 \ar[r]^\neg & s_3} 
	& \leadsto &
	\xymatrix{s_1 \ar@/_1pc/[rr]_\lor & s_2 \ar[r]^\neg & s_3
	} & \txt<5cm>{($s_2$ has a single outgoing transition)} \\
\hline
(4) & \xymatrix{s_1 \ar[r]^{\mu^k} & s_2}
	& \leadsto &
	\xymatrix{s_1 \ar[r]^\lor & s_2} 
	& \txt<5cm>{(decoding of $s_2$ does not contain $\holder{s_1}{k}$)} \\[3mm]
\hline
(5) & \xymatrix{s_1 \ar[r]^{\neg} & s_2}
	& \leadsto &
	\xymatrix{s_1 & s_2} 
	& \txt<5cm>{($s_2$ evaluates to $\true$)} \\[2mm]
\hline
(6) & 
	\xymatrix@C=.5pc@R=.8pc{
		& s_1 \ar[dl]_{\sigma_2} \ar[dr]^{\sigma_n} & \\
		s_2 & \ldots & s_n
	}
	& \leadsto &
	\xymatrix@C=.5pc@R=.8pc{
		& s_1 \ar[rr]^{\neg} & & \false \\
		s_2 & \ldots & s_n &
	} 
	& \txt<5cm>{($s_1$ evaluates to $\true$)} \\
\hline
(7) &
	\xymatrix{
		s_1 \ar[r]^{\sigma} & s_2
	}
	& \leadsto &
	\xymatrix{s_1 & s_2}
	& \txt<5cm>{($\sigma \neq \neg$ and $s_2$ evaluates to $\false$)} \\[4mm]
\hline
(8) &
	\xymatrix@C=.5pc@R=.8pc{
		& s_1 \ar[dl]_{\sigma_2} \ar[dr]^{\sigma_n} & \\
		s_2 & \ldots & s_n
	}
	& \leadsto &
	\xymatrix@C=.5pc@R=.8pc{
		& s_1 & & \\
		s_2 & \ldots & s_n &
	} 
	& \txt<5cm>{($s_1$ evaluates to $\false$)} \\
\hline
\end{array}
\]
\caption{Simplification rules applying to formula graphs}
\label{tab:simplification-rules}
\end{figure}

\subsection*{Elimination of $\lor$-transitions (1)}

	This rule allows transitions generated by synchronisation rules of the form $((\langle a\rangle, \vec{t}[i]), \lor)$ in the quotient formula network to be eliminated.
	This elimination can be achieved efficiently by applying reduction modulo $\tau^*.a$ equivalence~\cite{Fernandez-Mounier-91-a}, $\lor$-transitions being interpreted as internal ($\tau$) transitions.

\subsection*{Elimination of unguarded variables (2)}

	When combined with the previous rule, this rule allows unguarded variable occurrences to be eliminated.
	Indeed, an unguarded variable is characterized by a (possibly empty) sequence of $\lor$-transitions connecting the target and source of a $\mu$-transition.
	The elimination of this sequence of $\lor$ transitions then produces a self-looping transition labelled by $\mu$, which can be thereafter eliminated using the current rule.

\subsection*{Elimination of double-negations (3)}

	This rule can be used to simplify formulas of the form $\neg\neg\varphi$, which often occur in quotient formulas.
	For instance, a double-negation is introduced in the quotient of the formula $\neg\langle a\rangle\neg\varphi'$ with respect to an \LTS\ that offers an action synchronising with $a$ (thus having the modality disappear if the synchronisation is binary).

\subsection*{Elimination of $\mu$-transitions (4)}

	In this rule, the transition from $s_1$ to $s_2$ denotes the binder of a propositional variable $\holder{s_1}{k}$.
        If this variable does not occur free in the sub-formula denoted by state $s_2$, then the $\mu$-transition can be replaced by an $\lor$-transition, which can be subsequently eliminated using rule (1).
	Determining whether $\holder{s_1}{k}$ occurs free would require to decode the formula graph, which should be avoided in practice.
	For this reason, we only consider the following sufficient conditions, which can be checked in linear-time:

	\begin{itemize}
		\item $s_1$ and $s_2$ are not in the same strongly connected component (i.e., there is no path from $s_2$ to $s_1$), or

		\item $s_1$ is not the initial state and has a single predecessor $p$, and either $p$ has a single outgoing transition (which necessarily goes to $s_1$) and this transition is labelled by $\mu^{k'}$, or $p$ satisfies the same condition as $s_1$, recursively (this recursive condition is well-founded as long as it is applied to states reachable from the initial state)
	\end{itemize}

\subsection*{Evaluation of constant sub-formulas (5--8)}

	These four rules apply when some state denotes a sub-formula that evaluates to a constant in any context.
	This can be determined by using the following \BES, which implements partial evaluation of the formula.
	This \BES\ consists of blocks $T^k$ and $F^k$ ($k \in 0..n$) of respective signs $\mu$ and $\nu$, $n$ being the greatest block number in the formula graph.
	Blocks are ordered so that $k < k'$ implies $T^k$ (resp. $F^k$) is before $T^{k'}$ (resp. $F^{k'}$):
\[
\begin{array}{rrcl}
T^k: & \bigl\{~ T_{s}^k & =_{\mu} & \bigvee_{s \trans{\lor} s'} T_{s'}^k \lor
		  \bigvee_{s \trans{\neg} s'} F_{s'}^k \lor
		  \bigvee_{s \trans{\mu^{k'}} s'} T_{s'}^{k'} ~\bigr\}_{s \in \Sigma}
\\[2mm]
F^k: & \bigl\{~ F_{s}^k & =_{\nu} & \bigwedge_{s \trans{\lor} s'} F_{s'}^k \land 
		  \bigwedge_{s \trans{\langle\beta\rangle} s'} F_{s'}^k \land
		  \bigwedge_{s \trans{\neg} s'} T_{s'}^k \land
		  \bigwedge_{s \trans{\mu^{k'}} s'} F_{s'}^{k'} ~\bigr\}_{s \in \Sigma}
\end{array}
\]
	We consider only the variables reachable from $T_{s_0}^0$ or $F_{s_0}^0$, $s_0$ being the initial state of the formula graph.
	A state $s$ denotes $\true$ (resp. $\false$) if the Boolean variables $T_s^k$ (resp. $F_s^k$) evaluate to $\true$ in all (reachable) blocks $k$.
	Due to the presence of modalities, there may be states $s$ and blocks $k$ such that $T_s^k$ and $F_s^k$ are both false, indicating that the corresponding sub-formula is not constant.
	Intuitively, $T_s^k$ expresses that $s$ evaluates to $\true$ in block $k$ if one of its successors following a transition labelled by $\lor$ or $\mu^{k'}$ evaluates to $\true$, or one of its successors following a transition labelled by $\neg$ evaluates to $\false$.
	Variable $F_s^k$ expresses that state $s$ evaluates to $\false$ in block $k$ if all its successors following transitions labelled by $\lor$, $\mu^{k'}$, or modalities (by applying the identity $\langle a\rangle\false = \false$) evaluate to $\false$ and all its successors following transitions labelled by $\neg$ evaluate to $\true$.
	Regarding fix-point signs, observe that for the formula $\mu X^k.X^k$ (which is equivalent to the constant $\false$), $F_{\mu X^k.X^k}^k$ and $T_{\mu X^k.X^k}^k$ are defined respectively by the greatest fix-point equation $F_{\mu X^k.X^k}^k =_{\nu} F_{\mu X^k.X^k}^k$ and the least fix-point equation $T_{\mu X^k.X^k}^k =_{\mu} T_{\mu X^k.X^k}^k$.
This \BES\ has the solution $F_{\mu X^k.X^k}^k = \true, T_{\mu X^k.X^k}^k = \false$, reflecting the constant value false of $\mu X^k.X^k$ as expected.

	Repeated application of quotienting progressively eliminates modalities, until none of them remains in the formula graph, which then necessarily evaluates to a constant equal to the result of evaluating the formula on the whole network.

\subsection*{Sharing of equivalent sub-formulas}

	In addition to the above eight rules, reducing a formula graph modulo strong bisimulation does not change its decoding, modulo idempotence, renaming of propositional variables, and unification of equivalent variables defined in the same block.
	Strong bisimulation reduction can thus decrease the size of intermediate formula graphs.

\begin{exa}
\label{ex:simplifications}
	After applying the above simplifications to the formula graph of Example~\ref{ex:quotient-example} (page~\pageref{ex:quotient-example}), we obtain the (smaller) formula graph depicted in Figure~\ref{fig:formula-graphs} (c), page~\pageref{fig:formula-graphs}, which corresponds to the formula $(\langle a\rangle\true) \lor (\langle\alpha_a\rangle\true) \lor (\langle\alpha_b\rangle\langle a\rangle\true)$.
\end{exa}

\begin{exa}\label{ex:partial-evaluation}
	The graph corresponding to $\mu X^0.(\langle a\rangle\mu Y^0.\langle b\rangle X^0) \lor \langle c\rangle X^0$ reduces as expected to a deadlock state representing the constant $\false$ (left as an exercise).
\end{exa}

	Note that the simplification of a formula graph produces a formula graph.
	In particular, the parity of the number of occurrences of the label $\neg$ on paths leading to a $\mu^k$-transition is not changed by any rule, including rule~(3) which eliminates negations by pair.
	Also, the simplifications do not create new circuits and every $\mu^k$-transition eliminated by rule~(4) cannot be the first $\mu^k$-transition occurring on any circuit.

	All the simplifications that we propose in this paper correspond more or less to simplifications already proposed by Andersen~\cite{Andersen-95}, but we apply them directly on formula graphs instead of systems of $\mu$-calculus equations.
	For the interested reader, we review below the simplifications proposed by Andersen and detail how they map to our simplification rules:

\begin{itemize}
	\item {\em Reachability analysis\/} is included in our setting, due to our definition of the quotient on formulas (instead of systems of equations), which necessarily yields connected formulas (or formula graphs).
	In practice, reachability analysis is achieved using {\em on-the-fly\/} graph traversals, in particular on-the-fly generation of the \LTS\ corresponding to the quotient network.

	\item {\em Simple evaluation}, {\em constant propagation}, and {\em trivial equation elimination\/} are implemented by rules 5--8.
	The \BES\ that we have proposed for partial evaluation seems however slightly more general than Andersen's simplification rules, which do not seem to provide means to evaluate $X$ to $\false$ in the system of equations ``$X =_{\mu} \langle a\rangle Y \lor \langle c\rangle X, Y =_{\mu} \langle b\rangle X$'', whereas the corresponding formula (see Example~\ref{ex:partial-evaluation}) evaluates as expected to $\false$ in our setting.

	\item The approximation of {\em equivalence reduction\/} proposed by Andersen, which relies on a heuristic, is the same as our sharing of equivalent sub-formulas, implemented by strong bisimulation reduction.
	This can be seen easily as the definition of the heuristic in~\cite{Andersen-95} looks very similar to the definition of strong bisimulation on {\LTS}s.

	\item {\em Unguarded equations elimination\/} is implemented by the combination of rules 1--3.
\end{itemize}

\subsection*{About correctness of the simplifications}

The eight simplification rules preserve the semantics of the encoded formula.
We do not provide the formal proof of this statement, but we give the intuitions behind this result.
Intuitively, every rule defines a rather simple transformation on a set of equations.
Rule~(1) replaces the set $\{s_1 = s_2, s_2 = \psi\}$ by $\{s_1 = \psi, s_2 = \psi\}$, which is correct independently of the fix-point sign.
Rule~(2) replaces the equation $\{s_1 =_{\mu} s_1 \lor \psi\}$ by $\{s_1 =_{\mu} \psi\}$, which is a well-known transformation of the $\mu$-calculus.
Rule~(3) replaces $\{s_1 = \neg s_2 \lor \psi, s_2 = \neg s_3\}$ by $\{s_1 = s_3 \lor \psi, s_2 = \neg s_3\}$.
Rule~(4) reflects the fact that the fix-point sign of an equation does not influence the result of its resolution if the bound variable has no free occurrence in the set of equations.
Rules~(5) to~(8) express that any variable can be replaced by its solution.
At last, the sharing of equivalent formulas reflect that two variables can be merged if they are defined in the same block and if they have the same definition modulo variable names.
The correctness of a similar transformation has been proven in~\cite{Andersen-95}.

% -------------------------------------------------------------------------- %

\section{Simplification of Alternation-Free Formula Graphs}
\label{sec:alternation-free}

	Simplifications apply to $\mu$-calculus formulas of arbitrary alternation depth.
	We focus here on the alternation-free $\mu$-calculus fragment (\Lmui), which has a linear-time model checking complexity~\cite{Cleaveland-Steffen-93} and is therefore more suitable for scaling up to large {\LTS}s.
	We propose a variant of constant sub-formula evaluation specialised for alternation-free formulas, using alternation-free {\BES}s~\cite{Andersen-94}.

	Even in the case of alternation-free formulas, the above \BES\ is not alternation-free due to the cyclic dependency between $T^k$ and $F^k$, e.g., when evaluating sequences of $\neg$-transitions.
	In Figure~\ref{fig:alt-free-bes}, we propose a refinement of this \BES, which splits each variable $T_s^k$ of sign $\mu$ into two variables $T_s^{+k}$ of sign $\mu$ and $F_s^{-k}$ of sign $\nu$, which evaluate to true iff the sub-formula corresponding to state $s$ is preceded by an even (for $T_s^{+k}$) or odd (for $F_s^{-k}$) number of negations and evaluates to true.
	Variable $F_s^k$ is split similarly.
	This \BES\ is a generalisation, for formula graphs containing negations and modalities, of the \BES\ characterising the solution of alternation-free Boolean graphs outlined in~\cite{Mateescu-00-a}.

\begin{figure}[ht]
\[
\begin{array}{rl}
T^k: & \left\{\begin{array}{rcl}
T_s^{+k} & =_{\mu} & \bigvee_{s \trans{\lor} s'} T_{s'}^{+k} \lor
	  	     \bigvee_{s \trans{\neg} s'} T_{s'}^{-k} \lor
	  	     \bigvee_{s \trans{\mu^{k'}} s'} T_{s'}^{+k'} \\
T_s^{-k} & =_{\mu} & \bigwedge_{s \trans{\lor} s'} T_{s'}^{-k} \land
		     \bigwedge_{s \trans{\langle\beta\rangle} s'} T_{s'}^{-k} \land
		     \bigwedge_{s \trans{\neg} s'} T_{s'}^{+k} \land
		     \bigwedge_{s \trans{\mu^{k'}} s'} F_{s'}^{+k'}
\end{array}\right\}_{s \in \Sigma} \\[6mm]
F^k: & \left\{\begin{array}{rcl}
F_s^{+k} & =_{\nu} & \bigwedge_{s \trans{\lor} s'} F_{s'}^{+k} \land
		     \bigwedge_{s \trans{\langle\beta\rangle} s'} F_{s'}^{+k} \land
		     \bigwedge_{s \trans{\neg} s'} F_{s'}^{-k} \land
		     \bigwedge_{s \trans{\mu^{k'}} s'} F_{s'}^{+k'} \\
F_s^{-k} & =_{\nu} & \bigvee_{s \trans{\lor} s'} F_{s'}^{-k} \lor
	  	     \bigvee_{s \trans{\neg} s'} F_{s'}^{+k} \lor
	  	     \bigvee_{s \trans{\mu^{k'}} s'} T_{s'}^{+k'}
\end{array}\right\}_{s \in \Sigma}
\end{array}
\]\vspace{-4mm}
\caption{\BES\ for the evaluation of constant alternation-free formulas}
\label{fig:alt-free-bes}
\end{figure}

	\noindent For general formulas, this \BES\ is not alternation-free due to the cyclic dependencies between $T^k$ and $F^{k'}$, of different fix-point signs.
	Yet, for alternation-free block-labelled formulas, it is alternation-free, since each dependency from $T^k$ to $F^{k'}$ (or from $F^k$ to $T^{k'}$) always traverses a $\mu$-transition preceded by an odd number of negations, which switches to a different block number $k' > k$.

% -------------------------------------------------------------------------- %

\section{Handling fairness operators}
\label{sec:fairness}

	In the previous sections, we described a partial model checking procedure for the full modal $\mu$-calculus \Lmu, which we then specialised to the alternation-free fragment \Lmui.
	This fragment allows to express certain simple fairness operators, such as the fair reachability of actions (i.e., potential reachability by skipping cycles), originally proposed in the state-based setting~\cite{Queille-Sifakis-83}.
	The fair reachability of an action $a$ is expressed by the following \Lmui\ formula (where $\neg a$ denotes all actions except $a$), stating that as long as $a$ has not been encountered, it is still possible to reach it:
	$\nu X . (\mu Y . (\dia a \mond \true \vee \dia \true \mond Y) \wedge \brac \neg a \ket X)$.
	An equivalent, more concise, formulation of this property using the operators of \PDL~\cite{Fischer-Ladner-79} is $\brac (\neg a)^{*} \ket \dia \true^{*} . a \mond \true$.

	More elaborate fairness properties can be conveniently expressed by characterizing unfair cycles using the infinite looping operator $\Delta R$ of \PDLdelta~\cite{Streett-82}, which states the existence of an infinite transition sequence made by concatenation of subsequences that satisfy the regular expression $R$.
	The $\Delta R$ operator can be translated into the fix-point formula $\nu X . \dia R \mond X$, which can be further expanded into a plain $\mu$-calculus formula~\cite{Emerson-Lei-86}.
	This operator can encode the existence of accepting cycles in B\"uchi automata, and therefore it is able to capture \LTL\ properties; in fact, this operator brings significant expressive power to \PDL, making \PDLdelta\ more expressive than \CTLstar~\cite{Wolper-82}.
	When the regular expression $R$ contains Kleene star operators, the operator $\Delta R$ yields a formula of \Lmuii, the $\mu$-calculus fragment of alternation depth~2.
	Although this fragment has a quadratic worst-case model checking complexity~\cite{Emerson-Lei-86}, the $\Delta R$ operator can be checked on-the-fly in linear-time by formulating the problem as a \BES\ resolution and applying the \Acyc\ algorithm~\cite{Mateescu-Thivolle-08}.
	This algorithm generalizes the resolution algorithm A$_4$ for disjunctive \BES{s}~\cite{Mateescu-06-a} by enabling the detection of cycles in the underlying Boolean graphs that pass through marked Boolean variables, in a way similar to the detection of accepting cycles in B\"uchi automata.
	However, this does not yield a linear-time model checking for \LTL\ (resp. \CTLstar) because the translations from \LTL\ model checking problems to B\"uchi automata (resp. from \CTLstar\ formulas to \PDLdelta) are not succinct.

	We propose a way to evaluate the $\Delta R$ operator on a network of \LTSs\ using partial model checking, without developing the complex (and quadratic-time) machinery needed to evaluate general \Lmuii\ formulas.
	We rely instead on the approach proposed in~\cite{Mateescu-Thivolle-08}, which transforms the evaluation of $\Delta R$ into the resolution of an alternation-free \BES\ containing marked Boolean variables.
	We first illustrate this approach using an example of $\Delta R$ operator where $R$ contains star operators, and then we show its application in the partial model checking framework.

	Consider the formula $\Delta ((a|b)^{*}.c)$, which is equivalent to the  \Lmu\ formula $\nu X . \dia (a|b)^{*}.c \mond X$.
	The regular diamond modality can be further expanded by repeatedly applying the classical \PDL\ identities
	($\dia R_1 . R_2 \mond \varphi = \dia R_1 \mond \dia  R_2 \mond \varphi$,
	$\dia R_1 | R_2 \mond \varphi = \dia R_1 \mond \varphi \vee \dia R_2 \mond \varphi$, and
	$\dia R^{*} \mond \varphi = \mu Y . (\varphi \vee \dia R \mond Y)$)
	until all regular operators have been eliminated:
\[
\begin{array}{rcl}
\nu X . \dia (a|b)^{*}.c \mond X
& = & \nu X . \dia (a|b)^{*} \mond \dia c \mond X \\
& = & \nu X . \mu Y . (\dia c \mond X \vee \dia a|b \mond Y) \\
& = & \nu X . \mu Y . (\dia c \mond X \vee \dia a \mond Y \vee \dia b \mond Y)
\end{array}
\]
	The resulting \Lmuii\ formula can be written equivalently as a modal equation system containing two mutually recursive blocks with opposite fix-point signs:
\[
\{ X {=}_{\nu} Y \},
\{ Y {=}_{\mu} \dia c \mond X \vee \dia a \mond Y \vee \dia b \mond Y\}
\]
	The evaluation of variable $X$ on a state $s$ is reformulated as the resolution of the Boolean variable $X_s$ of the following \BES:
\[
{\textstyle
\{ X_s {=}_{\nu} Y_s \}_{s \in S},
\{ Y_s {=}_{\mu} \bigvee_{s \Arrow{c} s'} X_{s'} \vee \bigvee_{s \Arrow{a} s'} Y_{s'} \vee \bigvee_{s \Arrow{b} s'} Y_{s'} \}_{s \in S}
}
\]
	We observe that the $\nu$-block contains only singular equations, the $\mu$-block is disjunctive (i.e., all right-hand sides of equations contain only disjunctions), and does not contain $\true$ constants but possibly $\false$ constants (which correspond to empty disjunctions).
	This structure, which is guaranteed by construction for every \BES\ encoding the evaluation of a $\Delta R$ operator, enables to obtain a linear-time resolution procedure in the following way:
	(a) The $\nu$-block is merged into the $\mu$-block by changing the fix-point sign of its equations (this operation is abusive, since it changes the semantics of the \BES);
	(b) In the resulting $\mu$-block, the $X_s$ Boolean variables are marked (with the superscript $^@$) in order to retrieve the original semantics of the \BES\ during resolution.
	For the example considered, this procedure yields the following single-block \BES:
\[
{\textstyle
\{ X_s^@ {=}_{\mu} Y_s, Y_s {=}_{\mu} \bigvee_{s \Arrow{c} s'} X_{s'}^@ \vee \bigvee_{s \Arrow{a} s'} Y_{s'} \vee \bigvee_{s \Arrow{b} s'} Y_{s'} \}_{s \in S}
}
\]
	If the \LTS\ does not contain any infinite sequence belonging to the $\omega$-regular language $((a|b)^{*}.c)^{\omega}$, the initial formula $\Delta ((a|b)^{*}.c)$ evaluates to $\false$, which is also the result of evaluating variable $X_{s_0}$ in the $\mu$-block above.
	If there exists such an infinite sequence going out of the initial state $s_0$, the initial formula evaluates to $\true$, whereas variable $X_{s_0}$ in the $\mu$-block above does still evaluate to $\false$ (given the absence of $\true$ constants in this \BES).
	The existence of such an infinite sequence in the \LTS\ corresponds to a cycle in the Boolean graph associated to the \BES, which passes through some $X_s$ variable.
	Therefore, to retrieve the original semantics of the two-block \BES\, the resolution algorithm must mark the $X_s$ variables and detect whether one of these variables $X_s^@$ belongs to a cycle; if this is the case, then the variable is replaced by a $\true$ constant, which forces (by back-propagation through the disjunctive operators) the variable $X_{s_0}$ to evaluate to $\true$.

	This kind of resolution is carried out in linear-time by the $\Acyc$ algorithm~\cite{Mateescu-Thivolle-08}, based on a depth-first search of the Boolean graph with detection of cycles containing marked variables by computing the strongly connected components.
	This algorithm is robust w.r.t. repeated invocations, i.e., a sequence of calls has a cumulated linear-time complexity, which enables the evaluation of $\Delta R$ operators nested with (alternation-free) fix-point operators without losing the overall linear-time complexity in the size of the \BES.

	This evaluation procedure for $\Delta R$ operators can be applied in the partial model checking setting by abusively merging the two equation blocks into a single one, producing the formula graph in which the $X$ variable is marked (using an outgoing transition labeled by a special action $\mu @$), carrying out the projection steps, obtaining in the last step a modality-free formula graph corresponding to a \BES\ with marked variables, and solving this \BES\ using the \Acyc\ algorithm.
	During the projection steps, partial evaluation is carried out on the formula graph by using the same \BES\ as in Section~\ref{sec:alternation-free}, slightly extended to take into account the transitions labeled by $\mu_@^k$ corresponding to marked variables.
	Every $\mu$-block corresponding to a $\Delta R$ operator (with marked variables) is assigned a unique block number.
	Partial evaluation is carried out using algorithm $\Acyc$ every time a variable $Y$ belonging to such a block is encountered: if the algorithm detects a modality-free cycle containing a marked variable of that block, the variable $Y$ evaluates to $\true$.

\begin{figure}[htbp]
\begin{center}
\resizebox{\textwidth}{!}{
\input{fig_arobase.pstex_t}
}
\end{center}
\caption{Partial model checking of a fairness property on a network}
\label{fig:INFINITE-LOOPING}
\end{figure}

	Figure~\ref{fig:INFINITE-LOOPING} illustrates the partial model checking of a formula containing an infinite looping operator on a network representing a semaphore-based mutual exclusion protocol.
	The network $N = ((P_0, S, P_1), V)$, shown in Figure~\ref{fig:INFINITE-LOOPING}(a), consists of two processes $P_0$ and $P_1$ competing for a shared resource, and a semaphore $S$ guarding the access to the resource.
	Each process $P_i$ (for $i \in \{ 0, 1 \}$) cyclically executes the following sequence: first it performs its non-critical section ${\it ncs}_i$, then it requests the access to the resource by synchronising with the semaphore on ${\it req}_i$, then it accesses the resource during its critical section ${\it cs}_i$, and finally it releases the semaphore by synchronising on ${\it rel}_i$.
	The three processes interact via the following set of synchronisation vectors:
	\[
        \begin{array}{r@{~}l@{}l}
	V = \{ &
		  (({\it req}_0, {\it req}_0, \bullet), {\it req}_0), 
		(({\it rel}_0, {\it rel}_0, \bullet), {\it rel}_0), \\
		& ((\bullet, {\it req}_1, {\it req}_1), {\it req}_1), 
		((\bullet, {\it rel}_1, {\it rel}_1), {\it rel}_1), \\
		& (({\it ncs}_0, \bullet, \bullet), {\it ncs}_0), 
		(({\it cs}_0, \bullet, \bullet), {\it cs}_0), \\
		& ((\bullet, \bullet, {\it ncs}_1), {\it ncs}_1), 
		((\bullet, \bullet, {\it cs}_1), {\it cs}_1) & \}
	\end{array}
	\]
	The \PDLdelta\ formula checked on the network $N$ is
	$
		\brac {\it ncs}_0 \ket
		\Delta (
			(\neg {\it any}_0)^{*} .
			{\it ncs}_1 .
			(\neg {\it any}_0)^{*} .
			{\it cs}_1
		)
	$,
	stating that after $P_0$ executes its non-critical section, it may never access the shared resource because of a systematic overtaking by $P_1$ (the action formula $\neg {\it any}_0$ denotes the set of actions not executed by $P_0$, i.e., $\{ {\it ncs}_1, {\it req}_1, {\it cs}_1, {\it rel}_1 \}$).
	This formula can be expressed in \Lmu\ as
	$
		\nu X .
			\mu Y . (
			    \dia {\it ncs}_1 \mond
				\mu Z . (
				    \dia {\it cs}_1 \mond X
				    \vee
				    \dia \neg {\it any}_0 \mond Z
				)
			    \vee
			    \dia \neg {\it any}_0 \mond Y
			)
	$, or equivalently as the modal equation system
	$
	\{
	    U =_\mu \neg\dia {\it ncs}_0 \mond \neg X^@,
	    X^@ =_\mu Y,
	    Y =_\mu \dia {\it ncs}_1 \mond Z \vee \dia \neg {\it any}_0 \mond Y,
	    Z =_\mu \dia {\it cs}_1 \mond X^@ \vee \dia \neg {\it any}_0 \mond Z
	\}
	$,
	in which the equation defining $X^@$ has been abusively merged into the minimal fix-point block.
	The graph corresponding to this formula, where $X^@$ is marked by means of an outgoing transition labeled by $\mu_@^1$, is shown in Figure~\ref{fig:INFINITE-LOOPING}(b).

	At the last step of the partial model checking procedure (i.e., after quotienting w.r.t. processes $P_1$ and $S$), the formula graph obtained contains a modality-free cycle passing through $X$, indicated with thick arrows in Figure~\ref{fig:INFINITE-LOOPING}(c).
	This cycle is detected in linear-time by applying the simplification procedure, which invokes the \BES\ resolution algorithm \Acyc.
	We observe that the quotienting w.r.t. process $P_0$ was not necessary (and not done), since the presence of the cycle containing $X$ was detected as soon as processes $P_1$ and $S$ were taken into account.

% -------------------------------------------------------------------------- %

\section{Implementation}
\label{sec:implementation}

	We have implemented partial model checking of the alternation-free $\mu$-calculus extended with the $\Delta R$ fairness operator.
	We used \CADP, which provided much of what was needed:

\begin{itemize}
	\item Individual processes can be described in one of the numerous formats and languages available in \CADP: directly as {\LTS}s in, e.g. the \BCG\ file format\footnote{\url{http://cadp.inria.fr/man/bcg.html}}, or as high-level processes in the \LOTOS~\cite{ISO-8807}, \LOTOSNT~\cite{Champelovier-Clerc-Garavel-et-al-11} (a variant of \ELOTOS~\cite{ISO-15437}), or {\sc Fsp}~\cite{Magee-Kramer-06} languages.
	\CADP\ contains tools to generate {\LTS}s in the \BCG\ format automatically from those three languages.
	For the latter two, this is done via an automated generation of intermediate \LOTOS\ code using translators~\cite{Lang-Salaun-Herilier-et-al-10,Champelovier-Clerc-Garavel-et-al-11}.
	Other languages can easily be connected to \CADP\ using either the same approach (for instance a connection of the applied $\pi$-calculus~\cite{Mateescu-Salaun-10}), or through the \OPENCAESAR~\cite{Garavel-98} {\sc Api} of \CADP.

	\item Process compositions can be described in the \EXPOPEN~2.0 language~\cite{Lang-05}, which provides various parallel composition operators, such as synchronisation vectors~\cite{Arnold-89}, process algebra operators (\LOTOS, \CCS, \CSP, \mCRL), and the generalised parallel composition operator of \ELOTOS/\LOTOSNT~\cite{Garavel-Sighireanu-99}.
	It also provides generalised operators for hiding, renaming, and cutting labels based on a representation of label sets using regular expressions.
	The \EXPOPEN~2.0 tool compiles its input into a network of {\LTS}s.
	It then generates C code for representing the transition relation using the \OPENCAESAR\ interface~\cite{Garavel-98}, so that the \LTS\ can be either generated or traversed on-the-fly using various libraries.

	For partial model checking, the \EXPOPEN~2.0 tool has been slightly extended both to implement sub-network extraction and to generate the network representing the parallel composition between the formula graph and a chosen individual \LTS.

	\item Regular alternation-free $\mu$-calculus formulas (i.e., an extension of the alternation-free $\mu$-calculus with action formulas and regular expressions inside modalities to represent actions and sequences of actions) extended with the $\Delta R$ fairness operator can be handled by the \EVALUATOR\ on-the-fly model checker~\cite{Mateescu-Sighireanu-03,Mateescu-Thivolle-08}.
	Regular expressions inside modalities are eliminated by \EVALUATOR\ and replaced by ordinary fix-point formulas with mere action formulas inside the modalities.

	An option has been added for compiling the formula into a formula graph represented in the \BCG\ format.
	This option also takes as input the set of actions potentially occurring in the process composition (which can be obtained using \EXPOPEN~2.0), so that the action formulas can be replaced by finite sets of actions.

	\item Reductions modulo $\tau^*.a$ equivalence and strong bisimulation are achieved using respectively the {\sc Reductor} and {\sc Bcg\_Min} tools of \CADP, without any modification.
\end{itemize}

	\noindent Elimination of double-negations, of $\mu$-transitions, and evaluation of constant formulas (for \Lmui\ extended with the $\Delta R$ operator) have been implemented in a new prototype tool\footnote{This prototype tool, accompanied with a shell-script implementing partial model checking, a manual, and examples, can be downloaded at \url{http://convecs.inria.fr/software/pmc}. \CADP\ is required to be installed for the script and the prototype tool to run. \CADP\ licenses are free for academic users.} ($1,000$ lines of C code), which relies on the {\sc Caesar\_Solve} library~\cite{Mateescu-06-a} for solving alternation-free \BES\ (extended to handle fairness as explained in Section~\ref{sec:fairness}).
	Finally, the \LTS\ w.r.t. which the formula is quotiented at each step is selected automatically using the {\em smart\/} heuristic, described in~\cite{Crouzen-Lang-11}.

% -------------------------------------------------------------------------- %

\section{Experimentation}
\label{sec:experimentation}

	We have used partial model checking in two case studies, one in avionics addressing the verification of a communication protocol between a plane and the ground, based on {\sc Tftp} ({\em Trivial File Transfer Protocol\/})/{\sc Udp} ({\em User Datagram Protocol\/}) and the other one in hardware, addressing the verification of the bus arbitration protocol used in the {\sc Scsi}-2 standard.

\subsection{Trivial File Transfer Protocol/User Datagram Protocol}

	The {\sc Tftp}/{\sc Udp} case-study has been described by Garavel \& Thivolle in~\cite{Garavel-Thivolle-09}.
	In this section, we consider the same specifications and compare our new partial model checking approach with on-the-fly model checking.

	The system consists of two instances (A and B) of the {\sc Tftp} connected by {\sc Udp} using a {\sc Fifo} buffer.
	Since the state space of the specification is very large in the general case, Garavel \& Thivolle have defined five scenarios named $A$ to $E$, depending on whether each instance may write and/or read a file (see Table~\ref{tab:scenarios}).
        We have considered the same five scenarios in our study.
	All of them are specified in \LOTOS, as the parallel composition of eight processes named TFTP\_A, TFTP\_B, MEDIUM\_A, MEDIUM\_B, RCV\_A, RCV\_B, SND\_A, and SND\_B.
	The \LTSs\ corresponding to those eight processes are generated automatically from their \LOTOS\ specification using the {\sc Caesar} tool of \CADP.
	Their parallel composition is translated into a network of \LTSs\ using the \EXPOPEN\ tool of \CADP.
	Table~\ref{fig:lts-sizes} provides the sizes after reduction of the \LTSs\ corresponding to the eight processes for each scenario, as well as the size of their composition.

\begin{table}
\begin{center}
\begin{tabular}{|c||c|c||c|c|}
\hline
Scenario & \multicolumn{2}{c||}{TFTP A} & \multicolumn{2}{c|}{TFTP B} \\
\cline{2-5}
& read & write & read & write \\
\hline
\hline
A & & \tick & & \\
\hline
B & \tick & & & \\
\hline
C & & \tick & & \tick \\
\hline
D & \tick & & & \tick \\
\hline
E & \tick & & \tick & \\
\hline
\end{tabular}
\end{center}
\caption{The five scenarios of the TFTP/UDP case study}
\label{tab:scenarios}
\end{table}

\begin{table}
\begin{scriptsize}
\begin{center}
\begin{tabular}{|l|r|r|r|r|r|r|r|r|r|r|}
\cline{2-11}
\multicolumn{1}{c|}{} & \multicolumn{2}{c|}{Scenario A} & \multicolumn{2}{c|}{Scenario B} & \multicolumn{2}{c|}{Scenario C} & \multicolumn{2}{c|}{Scenario D} & \multicolumn{2}{c|}{Scenario E} \\
\cline{2-11}
\multicolumn{1}{c|}{} & \multicolumn{1}{c|}{States} & \multicolumn{1}{c|}{Trans.} & \multicolumn{1}{c|}{States} & \multicolumn{1}{c|}{Trans.} & \multicolumn{1}{c|}{States} & \multicolumn{1}{c|}{Trans.} & \multicolumn{1}{c|}{States} & \multicolumn{1}{c|}{Trans.} & \multicolumn{1}{c|}{States} & \multicolumn{1}{c|}{Trans.} \\
\hline
TFTP\_A & 704 & 4,542 & 719 & 4,610 & 704 & 4,542 & 719 & 4,610 & 719 & 4,610 \\
\hline
TFTP\_B & 504 & 3,421 & 504 & 3,421 & 1,058 & 7,164 & 1,058 & 7,164 & 1,058 & 7,164 \\
\hline
MEDIUM\_\{A,B\} & 801 & 5,440 & 801 & 5,440 & 801 & 5,440 & 801 & 5,440 & 801 & 5,440 \\
\hline
SND\_A, RCV\_B & 1 & 4 & 1 & 4 & 1 & 7 & 1 & 5 & 1 & 6 \\
\hline
SND\_B, RCV\_A & 1 & 4 & 1 & 3 & 1 & 7 & 1 & 6 & 1 & 6 \\
\hline
\hline
Product ($\times 10^3$) & $1,963$ & $8,527$ & 867 & $3,737$ & $35,024$ & $151,810$ & $40,856$ & $189,068$ & $19,436$ & $83,921$ \\
\hline
\end{tabular}
\end{center}
\end{scriptsize}
\caption{Individual \LTS\ sizes (in states and transitions) and product \LTS\ size (in kilostates and kilotransitions) for each scenario}
\label{fig:lts-sizes}
\end{table}

	We considered the (alternation-free) $\mu$-calculus (branching-time) properties named $A01$ to $A28$, studied in~\cite{Garavel-Thivolle-09}, as well as an additional alternation-2 fairness property $A29$ not checked in~\cite{Garavel-Thivolle-09}.
	We checked all properties both using the well-established on-the-fly model checker \EVALUATOR~\cite{Mateescu-Sighireanu-03,Mateescu-Thivolle-08} of \CADP\ and using the partial model checking approach described in this paper.
	These experiments were done on a $64$-bit computer with $148$ gigabytes of memory.

	The results summarized in Table~\ref{fig:experiments-memory} give, for each scenario and each property, the peak of memory in megabytes (MB) used by on-the-fly model checking (column fly) and partial model checking (column pmc).
	Some properties being irrelevant to some scenarios (e.g., they concern a read or write operation absent in the corresponding scenario), they have not been checked, which explains the shaded cells.
	The symbol ``$\star$'' corresponds to verifications that have been stopped because they took too long and used too much memory.
	The execution times are given in Table~\ref{fig:experiments-time}.
	Note that the major part of time and memory are used by formula simplifications, as compared to the rather low complexity of the synchronous product operation used for quotienting.

\newcommand{\fly}{\parbox{.9cm}{\centering fly}}
\newcommand{\pmc}{\parbox{.9cm}{\centering pmc}}
\newcommand{\ec}[1]{\multicolumn{2}{c#1}{\cellcolor{lightgray}}}
\begin{table}[t]
\begin{scriptsize}
\[
\begin{array}{|l||r|r||r|r||r|r||r|r||r|r|}
\cline{2-11}
\multicolumn{1}{c|}{}	& \multicolumn{2}{c||}{\mbox{Scenario}\ A}
			& \multicolumn{2}{c||}{\mbox{Scenario}\ B}
			& \multicolumn{2}{c||}{\mbox{Scenario}\ C}
			& \multicolumn{2}{c||}{\mbox{Scenario}\ D}
			& \multicolumn{2}{c|}{\mbox{Scenario}\ E} \\
\hline
\multicolumn{1}{|c||}{\mbox{Prop}}
			& \multicolumn{1}{c|}{\fly} & \pmc
			& \multicolumn{1}{c|}{\fly} & \pmc
			& \multicolumn{1}{c|}{\fly} & \pmc
			& \multicolumn{1}{c|}{\fly} & \pmc
			& \multicolumn{1}{c|}{\fly} & \pmc \\
\hline
\hline
A01  & 199 & 6 &  89 &  6 & 2,947 &    24 & 3,351 &    27 & 1,530 &    23 \\
\hline
A02  & 207 & 6 &  93 &  6 & 3,156 &    25 & 3,631 &    28 & 1,612 &    10 \\
\hline
A03  & 182 & 6 &  80 &  6 & 2,737 &     6 & 3,162 &     6 & 1,386 &     6 \\
\hline
A04  & 199 & 6 &  89 &  6 & 2,947 &     6 & 3,351 &    29 & 1,530 &     7 \\
\hline
A05  &  10 & 6 &   7 &  6 &     7 &     6 &     7 &     6 &    10 &    10 \\
\hline
A06  & 187 & 6 &  85 &  6 & 2,808 &     6 & 3,249 &     7 & 1,428 &     6 \\
\hline
A07  & 187 & 6 &  85 &  6 & 2,808 &     6 & 3,249 &     6 & 1,428 &     6 \\
\hline
A08  & 186 & 6 &  80 &  6 & 2,745 &     6 & 3,170 &     6 & 1,390 &     6 \\
\hline
A09a & \ec{||} &  \ec{||} &       \ec{||} & 3,290 &    28 & 1,488 &     6 \\
\hline
A09b & \ec{||} &  \ec{||} & 2,955 &     6 &       \ec{||} &        \ec{|} \\
\hline
A10  & \ec{||} &  \ec{||} & 3,354 &     6 &       \ec{||} & 1,674 &     6 \\
\hline
A11  & \ec{||} &  \ec{||} & 3,206 &     6 & 4,444 &     7 & 1,711 &     6 \\
\hline
A12  & \ec{||} &  \ec{||} &   620 & \star &   133 & \star &   101 & \star \\
\hline
A13  & \ec{||} &  \ec{||} &       \ec{||} & 4,499 & \star & 2,094 & \star \\
\hline
A14  & 267 & 6 &  \ec{||} & 3,988 &    23 &       \ec{||} & 2,107 &    15 \\
\hline
A15  & \ec{||} & 118 & 15 &   521 & \star &   156 & \star & 1,524 &    59 \\
\hline
A16  & \ec{||} &  \ec{||} &       \ec{||} &       \ec{||} &   186 &     8 \\
\hline
A17  & \ec{||} &  \ec{||} &   667 & \star &   569 & \star &        \ec{|} \\
\hline
A18  & \ec{||} &  85 &  6 &   476 &    11 &   255 &     6 & 1,391 &     6 \\
\hline
A19  & \ec{||} & 207 &  6 & 6,352 &    90 & 8,753 &    13 & 3,104 &    55 \\
\hline
A20  &  31 & 9 &  \ec{||} &   837 &    21 &       \ec{||} &   261 &    25 \\
\hline
A21  & 374 & 6 &  \ec{||} & 4,958 &    25 &       \ec{||} & 2,817 &    25 \\
\hline
A22  & \ec{||} &  35 &  7 &       \ec{||} &   427 & 1,271 &   191 &   650 \\
\hline
A23  & \ec{||} & 170 &  6 &       \ec{||} & 6,909 &     9 & 3,039 &    40 \\
\hline
A24  &  41 & 9 &  \ec{||} &   427 & 1,786 &       \ec{||} &        \ec{|} \\
\hline
A25  & 391 & 6 &  \ec{||} & 5,480 &    40 &       \ec{||} &        \ec{|} \\
\hline
A26  & 195 & 6 &  \ec{||} & 2,857 &    15 &       \ec{||} & 1,477 &    10 \\
\hline
A27  & 228 & 6 &  \ec{||} & 3,534 &     6 &       \ec{||} & 1,871 &     6 \\
\hline
A28  & \ec{||} & 102 &  6 & 3,654 &    22 & 4,032 &     6 & 1,821 &     6 \\
\hline
A29  & 198 & 7 &  88 &  7 & 2,942 &     9 & 3,350 &     7 & 1,525 &     9 \\
\hline
\end{array}
\]
\end{scriptsize}
\caption{Experimental results for the {\sc Tftp}/{\sc Udp} case study: memory (in megabytes)}
\label{fig:experiments-memory}
\end{table}

\begin{table}[t]
\begin{scriptsize}
\[
\begin{array}{|l||r|r||r|r||r|r||r|r||r|r|}
\cline{2-11}
\multicolumn{1}{c|}{}	& \multicolumn{2}{c||}{\mbox{Scenario}\ A}
			& \multicolumn{2}{c||}{\mbox{Scenario}\ B}
			& \multicolumn{2}{c||}{\mbox{Scenario}\ C}
			& \multicolumn{2}{c||}{\mbox{Scenario}\ D}
			& \multicolumn{2}{c|}{\mbox{Scenario}\ E} \\
\hline
\multicolumn{1}{|c||}{\mbox{Prop}}
			& \multicolumn{1}{c|}{\fly} & \pmc
			& \multicolumn{1}{c|}{\fly} & \pmc
			& \multicolumn{1}{c|}{\fly} & \pmc
			& \multicolumn{1}{c|}{\fly} & \pmc
			& \multicolumn{1}{c|}{\fly} & \pmc \\
\hline
\hline
A01  & 28 & 2 & 10 & 3 & 1,324 & 3 & 1,590 & 2 & 772 & 3 \\
% true true true true true
\hline
A02  & 31 & 3 & 12 & 3 & 1,640 & 6 & 2,010 & 7 & 883 & 6 \\
% true true true true true
\hline
A03  & 22 & 1 & 8 & 1 & 1,210 & 1 & 1,365 & 1 & 668 & 1 \\
% true true true true true
\hline
A04  & 26 & 3 & 10 & 3 & 1,400 & 3 & 1,598 & 3 & 770 & 3 \\
% true true true true true
\hline
A05  & 1 & 5 & 1 & 5 & 1 & 5 & 1 & 5 & 1 & 5 \\
% true true true true true
\hline
A06  & 23 & 3 & 9 & 3 & 1,306 & 3 & 1,540 & 3 & 667 & 3 \\
% true true true true true
\hline
A07  & 23 & 3 & 9 & 3 & 1,299 & 3 & 1,687 & 3 & 674 & 3 \\
% true true true true true
\hline
A08  & 22 & 3 & 8 & 3 & 1,220 & 3 & 1,620 & 3 & 625 & 3 \\
% true true true true true
\hline
A09a & \ec{||}     & \ec{||}     & \ec{||}      & 1,679 & 7 & 695 & 3 \\
% - - - true true
\hline
A09b & \ec{||}     & \ec{||}     & 1,415 & 8 & \ec{||}      & \ec{||} \\
% - - true - -
\hline
A10  & \ec{||} & \ec{||} & 2,112 & 3 & \ec{||} & 929 & 3 \\
% - - true - true
\hline
A11  & \ec{||} & \ec{||} & 1,722 & 3 & 3,583 & 1 & 997 & 3 \\
% - - true true true
\hline
A12  & \ec{||} & \ec{||} & 76 & \star & 8 & \star & 6 & \star \\
% - - true true true 
\hline
A13  & \ec{||} & \ec{||} & \ec{||} & 3,297 & \star & 1,446 & \star \\
% - - - true true 
\hline
A14  & 54 & 3 & \ec{||} & 2,681 & 3 & \ec{||} & 1,443 & 3 \\
% true - true - true
\hline
A15  & \ec{||} & 11 & 5 & 55 & \star & 15 & \star & 705 & 7 \\
% - false true true false
\hline
A16  & \ec{||} & \ec{||} & \ec{||} & \ec{||} & 40 & 1 \\
% - - - - false
\hline
A17  & \ec{||} & \ec{||} & 315 & \star & 217 & \star & \ec{|} \\
% - - true true -
\hline
A18  & \ec{||} & 9 & 1 & 86 & 7 & 35 & 3 & 599 & 1 \\
% - false false false false
\hline
A19  & \ec{||} & 53 & 3 & 6,159 & 3 & 9,393 & 3 & 2,697 & 3 \\
% - true true true true
\hline
A20  & 1 & 3 & \ec{||} & 224 & 6 & \ec{||} & 39 & 6 \\
% false - false - false
\hline
A21  & 131 & 3 & \ec{||} & 4,004 & 3 & \ec{||} & 2,293 & 3 \\
% true - true - true
\hline
A22  & \ec{||} & 1 & 12 & \ec{||} & 147 & 2,712 & 43 & 1,007 \\
% - true - true true
\hline
A23  & \ec{||} & 39 & 3 & \ec{||} & 5,605 & 9 & 2,345 & 6 \\
% - true - true true
\hline
A24  & 1 & 13 & \ec{||} & 148 & 3,189 & \ec{||} & \ec{|} \\
% true - true - -
\hline
A25  & 133 & 3 & \ec{||} & 4,163 & 6 & \ec{||} & \ec{|} \\
% true - true - -
\hline
A26  & 25 & 3 & \ec{||} & 1,383 & 3 & \ec{||} & 687 & 3 \\
% true - true - true
\hline
A27  & 38 & 3 & \ec{||} & 2,323 & 3 & \ec{||} & 1,196 & 3 \\
% true - true - true
\hline
A28  & \ec{||} & 15 & 3 & 2,538 & 3 & 2,615 & 3 & 1,277 & 3 \\
% - true true true true
\hline
A29  & 26 & 2 & 11 & 2 & 1,524 & 6 & 1,738 & 3 & 700 & 5 \\
\hline
\end{array}
\]
\end{scriptsize}
\caption{Experimental results for the {\sc Tftp}/{\sc Udp} case study: time (in seconds)}
\label{fig:experiments-time}
\end{table}

	These results confirm that partial model checking may be much more efficient (up to 600 times less memory in this example) than on-the-fly model checking.
	This is particularly the case of some formulas of either form $\brac R \ket \false$ or $\dia R \mond \true$, where $R$ is a regular expression, which denote the absence, respectively the existence, of a sequence of transitions that matches $R$.
	The quotient evaluates to true (in the case of formulas of the form $\brac R \ket \false$) or false (in the case of formulas of the form $\dia R \mond \true$) before all individual {\LTS}s have been taken into account in the quotient, because it has been possible to determine that none of the paths possible in the parts of the system already taken into account in the quotient may yield a path satisfying $R$ in the global system.
	We illustrate this by giving details on the verification of formula $A09b$ on Scenario $C$.
	This formula has the form $\brac R \ket \false$ and evaluates to true after the partial model checking steps reported in the following table.

\begin{center}
\begin{scriptsize}
\begin{tabular}{|l|r|r|}
\hline
Step & States & Transitions \\
\hline\hline
Initial formula graph                     &    13 &     62 \\
\hline
Simplification \& reduction               &     7 &     56 \\
\hline
Quotient wrt. TFTP\_A                     &   125 &  1,964 \\
\hline
Simplification \& reduction               &    60 &  1,512 \\
\hline
Quotient wrt. TFTP\_B                     & 9,166 & 69,490 \\
\hline
Simplification \& reduction               & 5,308 & 50,799 \\
\hline
Quotient wrt. MEDIUM\_B (encodes $\true$) &     2 &      1 \\
\hline
\end{tabular}
\end{scriptsize}
\end{center}

	The fairness formula $A29$ is also evaluated efficiently using partial model checking.
	This formula is specified in \PDL\ as $\Delta\,(\true^* . A_1 . (\neg (A_1 \lor A_2))^* . A_3 . (\neg A_1)^* . A_2$) (or, in the {\sc Mcl} input language of \EVALUATOR, as $\dia \true^* . A_1 . (\neg (A_1 \lor A_2))^* . A_3 . (\neg A_1)^* . A_2 \mond @$ ) and denotes the existence of a cyclic sequence of transitions matching the regular expression $\true^* . A_1 . (\neg (A_1 \lor A_2))^* . A_3 . (\neg A_1)^* . A_2$, where $A_1, A_2$, and $A_3$ are particular actions.
	It evaluates to false on all scenarios.
	The first steps of partial model checking for this formula on Scenario $E$ are detailed in the following table.
 
\begin{center}
\begin{scriptsize}
\begin{tabular}{|l|r|r|}
\hline
Step & States & Transitions \\
\hline\hline
\hline
Initial formula graph                          &     19 &     151 \\
\hline
Simplification \& reduction                    &      7 &     139 \\
\hline
Quotient wrt. TFTP\_B                          &    903 &  20,388 \\
\hline
Simplification \& reduction                    &    896 &  20,099 \\
\hline
Quotient wrt. TFTP\_A                          & 26,369 & 197,480 \\
\hline
Simplification \& reduction (encodes $\false$) &      1 &       0 \\
\hline
\end{tabular}
\end{scriptsize}
\end{center}
	In a few other cases, partial model checking leads to combinatorial explosion (properties $A12$, $A13$, $A15$, and $A17$) while on-the-fly model checking performs efficiently.
	We illustrate this with the verification of formula $A12$ on scenario $C$.
	This formula has the form $\dia R \mond \true$ and evaluates to true.
	The first steps of partial model checking are detailed in the following table, in which we provide the time and memory used to complete each step.
	The reduction step includes both the pre-reduction modulo $\tau^*.a$ equivalence (i.e., elimination of $\tau$-transitions) and the reduction modulo strong bisimulation.
        Note that this may produce a graph that is not minimal in number of transitions, although always minimal in number of states.

\begin{center}
\begin{scriptsize}
\begin{tabular}{|l|r|r|r|r|}
\hline
Step & Time (s) & Memory (MB) & States & Transitions \\
\hline\hline
Initial formula graph   &     &     &          8 &          56 \\
\hline
Simplification          &   0 &   4 &          8 &          56 \\
\hline
Reduction               &   0 &  66 &          4 &          52 \\
\hline
Quotient wrt. TFTP\_A   &   0 &  66 &        210 &       5,687 \\
\hline
Simplification          &   0 &   4 &        136 &       3,665 \\
\hline
Reduction               &   0 &  66 &        134 &       3,587 \\
\hline
Quotient wrt. TFTP\_B   &   0 &  66 &     21,172 &     168,172 \\
\hline
Simplification          &   0 &   6 &     21,015 &     168,172 \\
\hline
Reduction               &   1 &  66 &     14,042 &     119,789 \\
\hline
Quotient wrt. MEDIUM\_B &  14 &  66 &  1,648,096 &  10,327,294 \\
\hline
Simplification          &  35 & 267 &  1,648,089 &  10,327,294 \\
\hline
Reduction               &  72 & 234 &  1,551,338 &  14,773,975 \\
\hline
Quotient wrt. MEDIUM\_A & 686 & 540 & 40,572,824 & 229,050,227 \\
\hline
\multicolumn{5}{|c|}{\ldots} \\
\hline
\end{tabular}
\end{scriptsize}
\end{center}

	This explosion seems inherent to the structure of the system and the formula, intermediate quotients needing to capture a large part of the behaviour before the truth value of the formula can be computed.
	This shows that both partial and on-the-fly model checking are complementary and worthy of being used concurrently.

\subsection{The SCSI-2 Bus Arbitration Protocol}

	This case-study has been described by Garavel \& Hermanns in~\cite{Garavel-Hermanns-02}.
	It was originally designed to illustrate the combination of functional verification and performance evaluation features of \CADP.
	In this section, we reuse the specification\footnote{A \CADP\ demo available on-line at {\tt ftp://ftp.inrialpes.fr/pub/vasy/demos/demo\_31}.} to compare on-the-fly verification of an alternation-2 fairness formula with its verification using partial model checking.

	The case-study represents a storage system (developed by Bull in the early 90's) consisting of up to eight {\em devices\/} (up to seven hard disks and a disk controller) connected by a bus (which enables eight connections) implementing the {\sc Scsi}-2 standard.
	Each device is assigned a unique {\sc Scsi}-number ranging between $0$ and $7$, the device assigned the highest number having highest priority when several devices are ready to access the bus.
	Each disk is represented by a process named DISK\_$n$, the controller by a process named CONTROLLER\_$n$, and each unused connection to the bus by a process named NO\_DEVICE\_$n$, $n$ corresponding to the assigned {\sc Scsi}-number.
	The controller may send randomly to any disk of number $n$ a message ``{\tt CMD\ !}$n$'' ({\em command\/}) indicating a transfer request (read/write a block of data from/to the disk).
	After processing this command, the disk sends back to the controller a message ``{\tt REC\ !}$n$'' ({\em reconnect\/}).

	We considered the alternation-2 fairness property expressing that after the controller (of number $c$) sends a data transfer request to disk number $n$ such that $n < c$, then for each disk of number $m$ such that $m > n$, there must exist a cyclic execution sequence matching the regular expression $(\neg {\tt REC\ !}n)^* \cdot {\tt CMD\ !}m \cdot (\neg {\tt REC\ !}n)^* \cdot {\tt REC\ !}m$, i.e., the processing of data transfer request with a disk that has not priority over the controller does not prevent other requests to be processed by disks of higher priority.

	In a first step, we considered two different configurations (named $A$ and $B$) of the storage system, each consisting of three disks, one controller  and four unused connections.
	In configuration $A$, the controller is assigned number $7$ and the disks are assigned numbers $0$ to $2$.
	In configuration $B$, the controller is assigned number $1$ and the disks are assigned numbers $0$, $2$, and $3$.
	In both configurations, the \LTS\ corresponding to the system has $56,168$ states and $154,748$ transitions\footnote{Actually, a third configuration $C$ is proposed in the on-line \CADP\ demo, with the controller assigned number $0$. We have not considered this configuration as the fairness formula is trivially true in this case, the controller yielding priority to all disks.}.

	Configuration $A$ satisfies the fairness property.
	On-the-fly model checking takes $1.67$ seconds and $66$ MB, whereas partial model checking takes $4$ minutes and $107$ MB.
	The largest intermediate formula graph has $489,983$ states and $4,336,623$ transitions.
	On the contrary, configuration $B$ violates the property.
	On-the-fly model checking takes $1.12$ seconds and $66$ MB, whereas partial model checking takes $19.16$ seconds and $66$ MB.
	The largest intermediate formula graph has $22,171$ states and $198,467$ transitions.

	The performance of partial model checking on configuration $B$ is interesting, because intermediate formula graphs always remain smaller than the product \LTS.
	To see how this scales up, we evaluated the property on larger configurations, still assigning number $1$ to the controller, but progessively replacing the unused connections by additional disks (up to 6 disks, the configuration with 7 disks being too large for model checking).
	The results are given in Table~\ref{fig:scsi-experiment}.
	Note that partial model checking scales well on this example as, for configurations with five disks and more, it terminates faster than the product \LTS\ generation.
	We summarize in the following table the sizes of intermediate formula graphs during the partial model checking of the configuration with 6 disks.

\begin{center}
\begin{scriptsize}
\begin{tabular}{|l|r|r|r|r|}
\hline
Step & Time (s) & Memory (MB) & States & Transitions \\
\hline\hline
Initial formula graph       &       &        &        109 &         360 \\
\hline
Simplification              &     0 &      4 &          9 &          28 \\
\hline
Reduction                   &     0 &     66 &          6 &          25 \\
\hline
Quotient wrt. CONTROLLER\_1 &   734 &  1,165 & 19,545,220 & 332,937,946 \\
\hline
Simplification              & 1,021 &  7,630 & 19,072,829 & 332,937,946 \\
\hline
Reduction                   & 1,807 &  7,483 & 12,400,293 & 326,265,410 \\
\hline
Quotient wrt. NO\_DEVICE\_6 &   489 &  1,472 & 12,400,293 & 320,065,265 \\
\hline
Simplification              &   801 &  7,234 & 12,400,293 & 320,065,265 \\
\hline
Reduction                   & 2,219 &  4,673 & 12,400,293 & 547,718,843 \\
\hline
Quotient wrt. DISK\_0       & 1,073 &  2,657 & 29,367,067 & 710,452,069 \\
\hline
Simplification              &   721 & 17,594 &  1,345,007 &  36,186,832 \\
\hline
Reduction                   &   145 &    479 &  1,285,959 &  36,127,784 \\
\hline
Quotient wrt. DISK\_7       &   145 &    297 &  3,101,185 &  51,160,987 \\
\hline
Simplification              &   271 &  1,129 &  3,101,177 &, 51,160,987 \\
\hline
Reduction                   &   285 &    744 &  3,101,169 &  51,160,979 \\
\hline
Quotient wrt. DISK\_5       &   125 &    283 &  7,124,779 &  78,762,466 \\
\hline
Simplification              &   389 &  1,765 &  7,124,771 &  78,762,466 \\
\hline
Reduction                   &   652 &  1,398 &  6,024,459 & 103,247,732 \\
\hline
Quotient wrt. DISK\_4       &   276 &    623 & 13,770,325 & 152,237,584 \\
\hline
Simplification              &   971 &  3,449 & 13,770,317 & 152,237,584 \\
\hline
Reduction                   & 1,717 &  2,632 & 12,201,825 & 223,819,978 \\
\hline
Quotient wrt. DISK\_3       &   680 &  1,330 & 27,557,019 & 290,881,082 \\
\hline
Simplification              & 1,721 &  6,667 & 27,557,011 & 290,881,082  \\
\hline
Reduction                   & 5,099 &  5,571 & 25,967,207 & 442,140,277 \\
\hline
Quotient wrt. DISK\_2       & 1,002 &  2,521 & 44,137,283 & 343,601,116 \\
\hline
Simplification              &   417 &  7,791 &          1 &           0 \\
\hline
\end{tabular}
\end{scriptsize}
\end{center}

\begin{table}[t]
\begin{scriptsize}
\[
\begin{array}{|l|c|c|c|c|}
\cline{2-5}
\multicolumn{1}{c|}{}	& \multicolumn{4}{c|}{\mbox{Number of disks}} \\
\cline{2-5}
\multicolumn{1}{c|}{}	& 3 & 4 & 5 & 6 \\
\hline
\mbox{DISK\_$n$ size (states)} & 768 & 768 & 768 & 768 \\
\hline
\mbox{DISK\_$n$ size (transitions)} & 5,119 & 9,215 & 17,407 & 33,791 \\
\hline
\mbox{CONTROLLER\_$n$ size (states)} & 4,617 & 53,217 & 583,929 & 6,200,145 \\
\hline
\mbox{CONTROLLER\_$n$ size (transitions)} & 32,373 & 630\,828 & 12,237,723 & 238,990,986 \\
\hline
\mbox{NO\_DEVICE\_$n$ size (states)} & 1 & 1 & 1 & 1 \\
\hline
\mbox{NO\_DEVICE\_$n$ size (transitions)} & 16 & 32 & 64 & 128 \\
\hline
\hline
\mbox{Product \LTS\ size (states)} & 56,168 & 1,384,021 & 32,003,282 & 708,174,559 \\
\hline
\mbox{Product \LTS\ size (transitions)} & 154,748 & 4,499,237 & 119,691,662 & 2,992,012,087 \\
\hline
\mbox{Generation time (seconds)} & 1 & 17 & 884 & 31,193 \\
\hline
\mbox{Memory peak (MB)} & 66 & 66 & 680 & 17,594 \\
\hline
\multicolumn{1}{c|}{} & \multicolumn{4}{c|}{\mbox{On-the-fly model checking}} \\
\hline
\mbox{Verification time (seconds)} & 1 & 17 & 1,273 & 47,532 \\
\hline
\mbox{Memory peak (MB)} & 66 & 95 & 1,705 & 39,236 \\
\hline
\multicolumn{1}{c|}{} & \multicolumn{4}{c|}{\mbox{Partial model checking}} \\
\hline
\mbox{Verification time (seconds)} & 19 & 61 & 759 & 24,276 \\
\hline
\mbox{Memory peak (MB)} & 66 & 66 & 1,007 & 16,239 \\
\hline
\mbox{Largest formula graph (states)} & 22,171 & 253,723 & 2,773,147 & 29,367,067 \\
\hline
\mbox{Largest formula graph (transitions)} & 198,467 & 3,023,449 & 45,639,547 & 710,452,069 \\
\hline
\end{array}
\]
\end{scriptsize}
\caption{Experimental results for {\sc Scsi}-2 bus arbitration, 3 to 6 disks (configuration $B$)}
\label{fig:scsi-experiment}
\end{table}
%

% -------------------------------------------------------------------------- %

\section{Conclusion}
\label{sec:conclusion}

	The original contributions of this paper are the following:

\begin{enumerate}
	\item Partial model checking has been generalised to the network model, which subsumes many parallel composition operators.

	\item An efficient implementation of quotienting with respect to an individual \LTS\ has been proposed, using a synchronous product between this \LTS\ and a graph representation of the formula.
	A key is the representation of the formula in a disjunctive form (using negations), which turns every node of the formula graph into an {\em or-node}.

	\item An efficient implementation of formula simplifications has also been proposed, using a combination of existing algorithms (such as reductions modulo equivalence relations), simple transformations, and traversals of the formula graph using a \BES\ solver.
	Using a graph equivalence relation to simplify the formula was already proposed in~\cite{Basu-Ramakrishnan-03}, where the formula was translated into an {\em and-or-graph\/} and then reduced modulo strong bisimulation.
	We use a weaker relation ($\tau^*.a$ equivalence) that enables more reduction of the formula graph, and we apply it directly on simple {\LTS}s, thus allowing efficient \LTS\ reduction tools to be used without any modification.
	Our simplifications integrate smoothly in the approach, both quotienting and simplifications applying to the same graph representation, without encoding and decoding formulas back and forth.

	\item A specialisation to the case of alternation-free formulas (using alternation-free \BES) extended with the alternation-2 $\Delta R$ operator of \PDLdelta\ has also been proposed, and experiments have been conducted, showing that partial model checking may result in much better performance than complementary approaches, such as on-the-fly model checking.
	Only small software developments were required, thanks to the wealth of functionalities available in \CADP.
        The approach would be also applicable to formulas of arbitrary alternation depth using a solver for \BES\ of arbitrary alternation depth.
\end{enumerate}

	\noindent The implementation of quotienting as a synchronous product opens the way for combining partial model checking with techniques originating from compositional model generation, such as (compositional) $\tau$-confluence reduction~\cite{Lang-Mateescu-09,Mateescu-05,Pace-Lang-Mateescu-03}, or restriction using interface constraints following the approach developed in~\cite{Graf-Steffen-90} and refined in~\cite{Garavel-Lang-01,Krimm-Mounier-97,Lang-06}.
	Note also that partial model checking and compositional model generation are complementary.
	Although it is difficult in general to know which of them will be most efficient, a reasonable methodology is to try compositional model generation first (because one then obtains a single model on which all formulas of interest can be evaluated).
        In case of failure, partial model checking can then be used for each formula.

\IGNORE{
	The perspectives of this work are multiple: First, this approach should be extended to address the generation of verification diagnostics, to help the verifier understand truth values of formulas.
	Second, strategies should be designed to improve the performance of the approach, to answer questions such as: With respect to which individual \LTS\ in a network should a formula be quotiented first? In which order should the simplifications be applied? etc.

Perspectives : trouver des algorithmes plus efficaces en m\'emoire pour rendre l'approche compositionnelle plus efficace.
G\'en\'eraliser l'approche pour les logiques avec donn\'ees (type MCL).
Utiliser les r\'eseaux de stations de travail pour lancer les t\^aches de v\'erification en parall\`ele.
Support dans \SVL~\cite{Garavel-Lang-01}
}

% -------------------------------------------------------------------------- %

\subsection*{Acknowledgements}

	The authors warmly thank Hubert Garavel, Wendelin Serwe, and Damien Thivolle for providing the sources of case-studies.
	They thank the past and present developers of \CADP, without which this work would not have been possible.
	They also thank the anonymous referees, whose remarks greatly helped to improve this paper.

% -------------------------------------------------------------------------- %

% \bibliographystyle{cortex_plain}
% \bibliography{bibl_lotos,bibl_xtl,bibl_aux}

% -------------------------------------------------------------------------- %

% \newpage
% \appendix
% \setcounter{page}{1}
% \input{proofs}
% \input{formulas}

\end{document}